# Mechanistic discovery in grain boundary structure and deformation driven by agentic AI

Yazhuo Liu[a,†], Kunqing Ding[a,†], Xiaoguo Gong[b], Zifeng Wang[a], Yichen Yang[a], Yifeng Che[a], Wen Chen[d,e], Josh Kacher[c], Olivier Pierron[a], Shigenobu Ogata[b], Ting Zhu[a,c,*]

[a]Woodruff School of Mechanical Engineering, Georgia Institute of Technology, Atlanta, GA, 30332, USA

[b]Department of Mechanical Science and Bioengineering, Graduate School of Engineering Science, The University of Osaka, Osaka, 560-8531, Japan

[c]School of Materials Science and Engineering, Georgia Institute of Technology, Atlanta, GA, 30332, USA

[d]Department of Aerospace and Mechanical Engineering, University of Southern California, Los Angeles, CA, 90089, USA

[e]Mork Family Department of Chemical Engineering and Materials Science, University of Southern California, Los Angeles, CA, 90089, USA

## Abstract

Agentic artificial intelligence (AI) systems hold immense potential for automating complex workflows in materials discovery and design, but extending these capabilities to domain-intensive mechanistic studies remains challenging. General grain boundaries (GBs) exemplify this challenge, as their structural complexity complicates experimental characterization and atomistic modeling. Here, we present GBClaw, an agentic AI framework for investigating the structure and deformation of general GBs, which are common in real polycrystals yet remain understudied. By integrating large language models into a modular architecture, GBClaw effectively combines systematic simulations with targeted mechanistic analysis through iterative workflow refinement and dynamic replanning. Applied to 14 general asymmetric tilt GBs in nickel, the framework characterized interfacial defect structures and deformation modes across these incommensurate boundaries. It established a general approach to identify previously unobserved, widely spaced secondary GB misfit dislocations (SGBMDs) alongside closely spaced primary GB misfit dislocations (PGBMDs) in simulated atomic configurations and experimental images. Comparisons of GB sliding with and without SGBMDs revealed substantial differences in shear strength, underscoring the critical role of misfit-dislocation hierarchies in sliding resistance. Researcher-guided follow-up calculations further elucidated the structural origins of shear asymmetry and an anomalous shear response. This work demonstrates that combining agentic AI with domain expertise bridges high-throughput computational screening and deep mechanistic discovery at complex material interfaces.

† These authors contributed equally to this work.

* Corresponding author: ting.zhu@me.gatech.edu

## Introduction

Agentic artificial intelligence (AI) systems offer new ways to automate workflows and support discovery[1,2]. These systems integrate large language models (LLMs) with a harness comprising prompts, tools, middleware, and other components that enable iterative reasoning and task execution. Multiple agents can coordinate specialized capabilities to carry out complex workflows involving many interdependent steps. Although applications in materials research increasingly encompass information retrieval, computation, and data analysis[3,4], extending these capabilities to open-ended mechanistic investigations remains challenging. Such investigations require domain expertise to define research objectives, assess physical validity, resolve ambiguities, and select informative follow-up calculations[5,6]. Integrating researchers into the agentic workflow provides an effective means for addressing these requirements[5,7].

Grain boundary (GB) research presents a demanding application for this approach. GBs separate adjacent grains in polycrystalline materials and fundamentally dictate bulk mechanical behavior, including yield strength, tensile ductility, creep resistance, and fracture toughness[8–10]. Although general GBs constitute many real-world interfaces, they remain understudied because their structural complexity complicates experimental and modeling research. The macroscopic geometry of a planar GB possesses five degrees of freedom, and each geometry can admit multiple metastable atomic configurations[11–13]. Characterizing this structural diversity and its mechanical consequences[14,15] requires both consistent computational procedures across many interfacial configurations and adaptive investigations guided by intermediate results. A master agent (i.e., an orchestrator) can coordinate these activities by translating research objectives into tasks for specialized agents that use validated tools, generate custom code, or interface with scientific software[6,16]. Domain experts contribute by evaluating the results and identifying physical questions that require further investigation.

Here, we introduce GBClaw, an agentic framework built on the open-source LangChain platform[17] for automated GB simulations with researcher-guided mechanistic analysis. We demonstrate its capabilities on general asymmetric tilt GBs that lack an exact coincidence site lattice[18] (i.e., non-CSL). Such incommensurate interfaces[12,19,20] are common in polycrystalline materials, yet they remain insufficiently studied. GBClaw characterizes the atomic structures, shear strengths, and deformation pathways of these boundaries, while establishing automated procedures to identify interfacial misfit defects. The framework flags anomalous mechanical responses and dynamically refines workflows in response to researcher queries, conducting targeted simulations that uncover underlying physical mechanisms. These applications demonstrate how a human-in-the-loop agentic framework bridges systematic computational screening and a deep mechanistic understanding of complex material interfaces.

## GBClaw: Architecture and workflow

### Architecture

GB investigations combine sequential dependence with opportunities for parallel execution. Downstream tasks often depend on earlier results, whereas independent cases within a stage can be processed concurrently. Coordinating these activities requires task planning, information exchange, and persistent tracking of the investigation state. Leveraging the strong reasoning and analytical capabilities of LLMs, GBClaw supports these functions through a modular architecture consisting of an orchestrator, specialized agents, domain-specific tools, computational interfaces, and a file-based hierarchical memory system (Fig. 1). Researchers interact directly with the orchestrator, supported by the planning agent, to define research objectives, review intermediate findings, and refine subsequent investigations. The orchestrator delegates tasks, integrates the results, and produces a research report. Each specialized agent is routed to an LLM through an application programming interface (API) and operates with a bounded context, tailored prompt instructions, and designated tools. Separating coordination from execution limits the context required by each agent and supports concurrent processing of independent cases.

GBClaw includes five specialized agents with complementary roles (Fig. 1). The *planning agent* uses the scientific knowledge encoded in LLMs[7] to translate research objectives into specific questions, simulation parameters, and executable plans. The *structure agent* determines grain orientations, symmetry-consistent pairs of terminating planes, and structural periodicities needed for bicrystal construction. It supports both exact CSL geometries[21] and near-CSL constructions that accommodate departures from exact coincidence[22]. The *LAMMPS agent* constructs bicrystals and performs structural relaxation, GB energy calculation, shear simulations, and minimum-energy-path (MEP) searches using the nudged elastic band (NEB) method[23,24]. The *analysis agent* links shear stress responses to atomic displacements, structural changes, and GB transition pathways. It also identifies stress-drop events and generates visualizations to support mechanistic interpretation. The *coding agent* generates, executes and refines custom scripts for computational and data-processing tasks not covered by predefined tools[6,25]. The agents access LAMMPS[26],

OVITO[27], and ASE[28] through a unified Python integration layer. This modular design supports the addition of agents with further scientific capabilities.

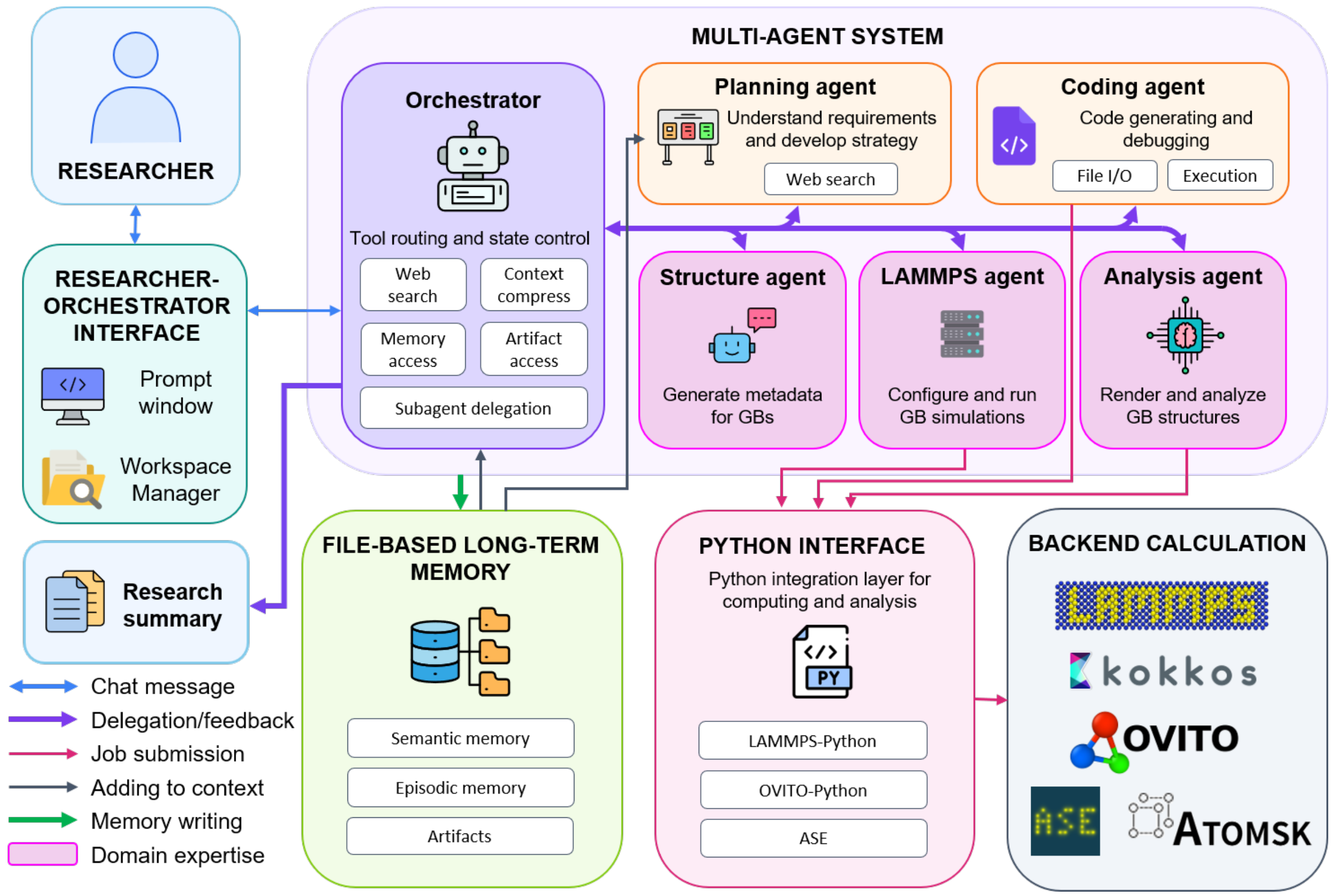

**Fig. 1 | Agentic architecture of GBClaw for researcher-guided grain boundary research.** The orchestrator serves as the central coordination hub, interfacing the researcher with five specialized agents, domain-specific scientific software tools, and a file-based hierarchical memory system. The modular architecture enables automated task decomposition and delegation, parallel simulation execution, structured information exchange, and interactive, human-in-the-loop workflow refinement.

GBClaw maintains continuity across tasks through structured information exchange. Each delegated task receives the context for execution, including established findings and references to previous simulations and computational artifacts. Artifacts include generated files, datasets, simulation outputs, code, and analysis results. When additional information is needed, specialized agents retrieve relevant records from project memory before invoking computational tools. On completion, each agent reports its status, findings, and artifact references to the orchestrator, which updates the project state and uses these results to guide subsequent tasks and researcher review. This approach provides focused task context while preserving access to the broader investigation. Because scientific interpretation depends on the model, computational conditions, and analysis procedure as well as final output, GBClaw retains these through hierarchical memory and context management as discussed next.

GBClaw employs a hierarchical memory mechanism. Inspired by the CoALA framework of working, episodic, semantic and procedural memories[29], GBClaw organizes information into five agent-facing levels (Supplementary Table S1). (i) Domain knowledge encodes workflows and task rules within system instructions and tools. (ii) The project summary tracks top-level objectives, key findings and conclusions for ongoing coordination. (iii) The conversation state records recent agent exchanges, decisions and open tasks. (iv) Persistent research records store durable findings, run summaries, and output references. (v) The artifact level contains underlying configurations, trajectories, logs, code and analysis outputs. Beyond these agent-facing levels, GBClaw maintains a separate archive containing complete interaction and tool execution histories for researcher review. Excluding this archive from routine agent retrieval reduces the risk of treating unverified conversational content as validated findings, while retaining records of how the investigation developed.

GBClaw manages active context through summary updates and conversation compaction. Long context can impair an LLM's retrieval and use of relevant information[30], so the system maintains concise project summaries and periodically condenses conversation histories. Before proposed research records are stored, provenance checks link their claims to supporting sources and simulation results. These checks support traceability but do not independently establish physical correctness. Project summaries incorporate new findings and simulation records, while conversation compression retains key decisions, pending tasks, and artifact references without deleting the underlying computational data. When more detail is needed, agents use shared identifiers to retrieve indexed records and exact run files. These persistent links also allow downstream simulation and analysis tools to access the original datasets directly (Supplementary Fig. S1). Such management connects scientific interpretation to its supporting models and outputs, thereby enabling researchers to review evidence and guide further investigation.

**Structural screening and targeted simulations**

General asymmetric, non-CSL tilt GBs are common in polycrystalline materials, yet they remain insufficiently understood. These boundaries provide an ideal platform for investigating how complex interfacial structures govern deformation. Experimental studies have demonstrated that GBs frequently feature low-index terminating planes on one or both sides[20,31–37], motivating our focus on GBs bounded by simple crystallographic terminations. In these non-CSL tilt GBs, the opposing surface lattices exhibit an irrational ratio of repeat distances along the sliding direction, rendering them inherently incommensurate[20,35]. This interfacial mismatch can be accommodated either by elastic strain or through a hierarchy of GB misfit dislocations in extended boundaries[11,34,38]. Under shear loading, these interfaces can undergo pure sliding (tangential translation only) or shear-coupled migration (tangential translation paired with normal boundary motion)[14,39]. Resolving these deformation mechanisms requires both high-throughput structural screening and detailed analysis of local atomic rearrangements during deformation.

GBClaw investigated flat, atomically sharp GBs in face-centered cubic (FCC) Ni bicrystals through systematic structural screening combined with researcher-guided shear simulations and analysis (Supplementary Fig. S2). ⟨110⟩ tilt GBs were selected because they are frequently observed in polycrystals, including in HRTEM images[36,37], thereby providing structural benchmarks for evaluating atomistic models. The detailed workflow is presented in Supplementary Fig. S3. Representative researcher-GBClaw interactions are shown in Supplementary Fig. S4 and the demonstration page (https://zhugroup.gatech.edu/gbclaw-demo). During screening, the structure agent used an LLM to identify near-Σ GBs containing low-index terminating planes and having misorientations close to those of reference Σ GBs[40,41]. Candidate boundaries were required to satisfy three criteria: (i) at least one terminating plane belongs to the {100}, {110}, or {111} family; (ii) the deviation of misorientation angle from the corresponding reference Σ GB does not exceed 5°; and (iii) the absolute values of all Miller indices of both terminating planes are less than 4. The selected boundaries were asymmetric and incommensurate. Their bicrystallographic parameters were then used to construct the corresponding bicrystal models, which the LAMMPS agent relaxed to characterize their interfacial defect structures.

Relaxation of extended boundaries produced a hierarchy of GB misfit dislocations[22]. Closely spaced, non-glissile primary GB misfit dislocations (PGBMDs) accommodate local mismatch, whereas more widely spaced glissile secondary GB misfit dislocations (SGBMDs) accommodate the residual mismatch. The shortest complete repeat of the PGBMD arrangement defines the primary periodicity $\boldsymbol{\lambda}$, while the characteristic spacing between SGBMDs defines the secondary periodicity $\boldsymbol{\Lambda}$ within a zeroth-order approximation[22]. General procedures are developed to identify SGBMDs and analyze their characters. Although SGBMD formation is usually energetically favorable in extended boundaries, models that exclude these defects accommodate the residual mismatch through elastic strain. The following sections present the screening results, compare the shear responses with and without SGBMDs, and describe the researcher-guided investigations of an anomalous shear response and shear asymmetry.

## Results

### GB sliding and shear strength

To investigate the shear behavior of asymmetric non-CSL GBs, the structure agent applied the screening criteria described above to general asymmetric ⟨110⟩ tilt GBs near reference Σ GBs. This screening yielded 12 distinct pairs of terminating planes, each corresponding to an asymmetric non-CSL GB (Supplementary Table S2). Some plane pairs were associated with multiple reference Σ GBs. For example, the {111}/{110} pair lies near both the Σ9 and Σ27 GBs. A given plane pair can also exhibit multiple crystallographically distinct misorientations. For example, the {111}/{331} pair includes variants at approximately 22° and 49°, whereas the {111}/{221} pair includes variants at approximately 16° and 55°. In addition, the initial screening near Σ33a identified only the 22° and 16° configurations. To enable direct comparisons between the variants, researcher guidance added the corresponding 49° and 55° configurations, increasing the final set to 14 distinct GB configurations.

Following selection of the asymmetric non-CSL GBs, the LAMMPS agent constructed and relaxed the corresponding Ni bicrystals, and the analysis agent examined their interfacial atomic structures. Interatomic interactions were modeled using the angular-dependent potential developed by Howells and Mishin[42], which reproduced the experimentally observed interfacial motif of the {111}/{331}(22°) GB[36]. This boundary exemplifies the hierarchy of PGBMDs and SGBMDs. Energy minimization of an extended boundary without periodic boundary conditions (PBCs) along the horizontal direction produced closely spaced PGBMDs with a Burgers vector of $\mathbf{b}_p$ = 1/2⟨110⟩{111} alongside more widely spaced SGBMDs whose Burgers vector $\mathbf{b}_s$ will be determined later. As shown in Fig. 2a, b, small and large red regions with high atomic volumetric strains near the {331} terminating plane identify the cores of PGBMDs and SGBMDs, respectively. The primary periodicity, defined by the repeating arrangement of PGBMDs, is $\boldsymbol{\lambda}$ = 5/2⟨112⟩{111}, referenced to the grain terminated by the {111} plane. The secondary periodicity $\boldsymbol{\Lambda}$ is the spacing between adjacent SGBMDs, corresponding to an integer multiple of $\boldsymbol{\lambda}$. The magnitude of $|\boldsymbol{\lambda}|$ = 2.16 nm is substantially smaller than that of $|\boldsymbol{\Lambda}|$ = 15.73 nm (Fig. 2b).

To distinguish macroscopically uniform GB shearing from sliding mediated by SGBMD glide, the LAMMPS agent constructed a compact periodic bicrystal model based on the primary periodicity λ. An interfacial region between neighboring SGBMDs in the extended GB (Fig. 2c) was isolated to build a supercell with PBCs within the GB plane (Fig. 2d). After energy minimization, the supercell retained its local PGBMD arrangement, while accommodating residual interfacial mismatch through elastic strain (Fig. 2c, d). These compact models enabled the evaluation of athermal shear strengths and deformation modes in the absence of SGBMDs under imposed periodic constraints.

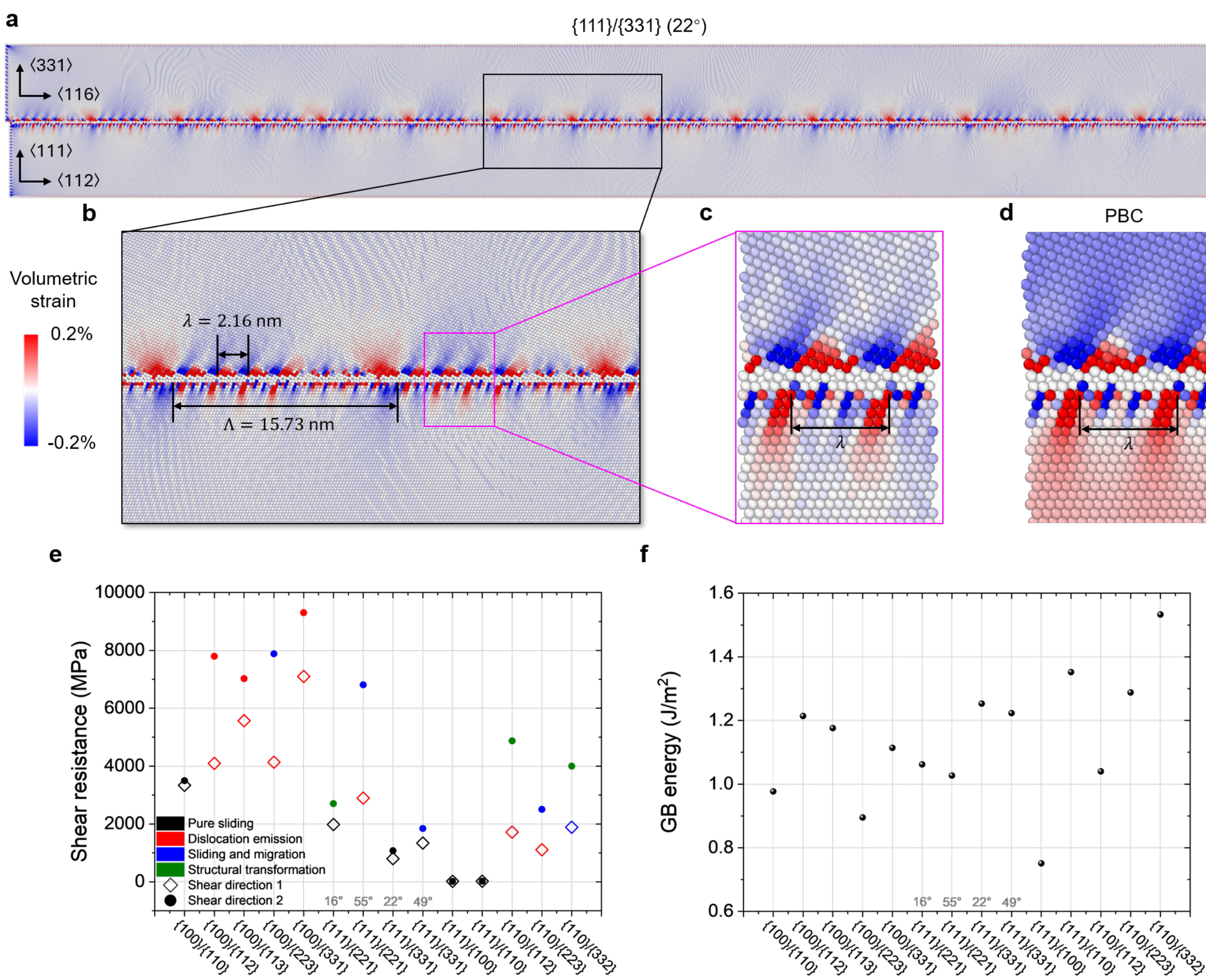

**Fig. 2 | Atomic structures, structural periodicities, and shear responses of selected asymmetric incommensurate GBs. a,** Extended GB in a Ni bicrystal after energy minimization, joining {331} and {111} terminating planes at a 22° misorientation angle, denoted as {111}/{331}(22°). **b,** Magnified view of the interface. Small and large red regions near the {331} terminating plane identify the cores of PGBMDs and SGBMDs, respectively. The primary period

($\lambda = 2.16$ nm) is the spacing between the repeating arrangement of PGBMDs, and the secondary period ($\Lambda = 15.73$ nm) is the spacing between adjacent SGBMDs. **c,** Magnified view of the boundary showing adjacent PGBMDs separated by one $\lambda$. **d,** Periodic supercell constructed from the structure in **c** and relaxed under PBCs to eliminate normal stress components for uniform sliding simulation. Atoms in **a**-**d** are colored according to atomic volumetric strain. **e,** Athermal shear strengths of the selected asymmetric incommensurate GBs loaded in opposite directions parallel to the GB plane and perpendicular to the $\langle 110 \rangle$ tilt axis. Directions 1 and 2 denote the lower and higher shear resistance, respectively. Symbol colors represent the observed deformation modes. **f,** Relaxed GB energies per unit area for the compact boundary models in **d**.

Using these compact models, the orchestrator coordinated multiple LAMMPS agents to parallelize shear simulations across all 14 GB configurations in opposing shearing directions (Fig. 2e). Horizontal shear displacement was applied incrementally, followed by energy minimization after each increment (Methods). The shear strength $\tau_0$, defined as the peak shear stress in the subsequent steady response, varied widely among these boundaries, ranging from tens of MPa to 9.3 GPa. Exceptionally low $\tau_0$ values of 23 MPa and 14 MPa for the $\{111\}/\{110\}$ and $\{111\}/\{100\}$ GBs suggest potential for low-friction sliding. In contrast, the high $\tau_0$ value of 3.5 GPa for the $\{100\}/\{110\}$ GB challenges its previously proposed suitability as a low-friction interface[43]. Pronounced stress drops accompanied diverse inelastic deformation modes, including pure sliding, shear-coupled migration, dislocation emission, and structural transformations that altered the initial boundary structure. Reversing the initial loading direction frequently changed the shear strength and, in some cases, the deformation mode. For instance, the $\{111\}/\{331\}(49°)$ GB exhibited pure sliding in one direction and shear-coupled migration in the reverse. These results demonstrate a pronounced direction-dependent asymmetry in shear responses, shifting the active deformation mechanism even within identical initial boundary configurations.

To examine the relationship between GB energy and shear resistance, the LAMMPS agent calculated the energies of all relaxed boundaries (Fig. 2f and Supplementary Note 1). Although higher-energy boundaries generally exhibited a trend toward lower shear resistance, this relationship was non-monotonic. For example, the $\{111\}/\{100\}$ GB had both the lowest relaxed GB energy and the lowest shear strength of 14 MPa. Furthermore, for most boundaries lacking structural symmetry about the interface normal, loading in opposite directions yielded different shear strengths and deformation modes despite identical GB energies. These findings demonstrate that static GB energy alone is insufficient to predict or characterize the shear responses of these general GBs, underscoring the need to directly analyze atomic rearrangements and deformation pathways under shear loading.

**Pure sliding**

Four GB configurations exhibited pure sliding under both shear directions: the {111}/{100}, {111}/{331}(22°), {100}/{110} and {111}/{110} GBs. To investigate their deformation pathways, the analysis agent correlated stress-drop events with corresponding atomic rearrangements (Fig. 3). Each GB contained at least one close-packed {111} or densely packed {100} terminating plane, whereas differences in the opposing terminating planes produced distinct interfacial registries. Specifically, the {110} termination exhibited atomic-scale corrugation, whereas the {331} termination comprised single-atomic-layer steps separated by short {111} facets. These geometric differences resulted in distinct local structural motifs within each interfacial repeat unit. Table 1 lists the primary periodicities $\lambda$ determined for all four GBs.

Comparisons of atomic configurations across selected sliding intervals, particularly immediately before and after stress drops, revealed the associated interfacial displacement fields (Fig. 3). Within a single structural repeat unit, atomic displacements were either distributed across multiple atomic columns or localized to individual columns oriented along the ⟨110⟩ tilt axis. Atoms within adjacent columns translated in different directions, indicating that macroscopically uniform grain translation was accommodated by locally nonuniform atomic shuffling. The spatial distribution of these rearrangements varied markedly among the four pure-sliding interfaces. While both the {111}/{100} and {111}/{110} GBs exhibited low shear strengths (Fig. 3a, b), displacements were distributed more uniformly in the {111}/{100} GB than in the {111}/{110} GB. The {111}/{331}(22°) and {100}/{110} GBs exhibited substantially higher shear strengths, but also differed in their local displacement modes. In the {111}/{331}(22°) GB, displacements were distributed across all atomic columns with no single column dominating the response. In contrast, each stress-drop event in the {100}/{110} GB involved pronounced shuffling localized within a single column.

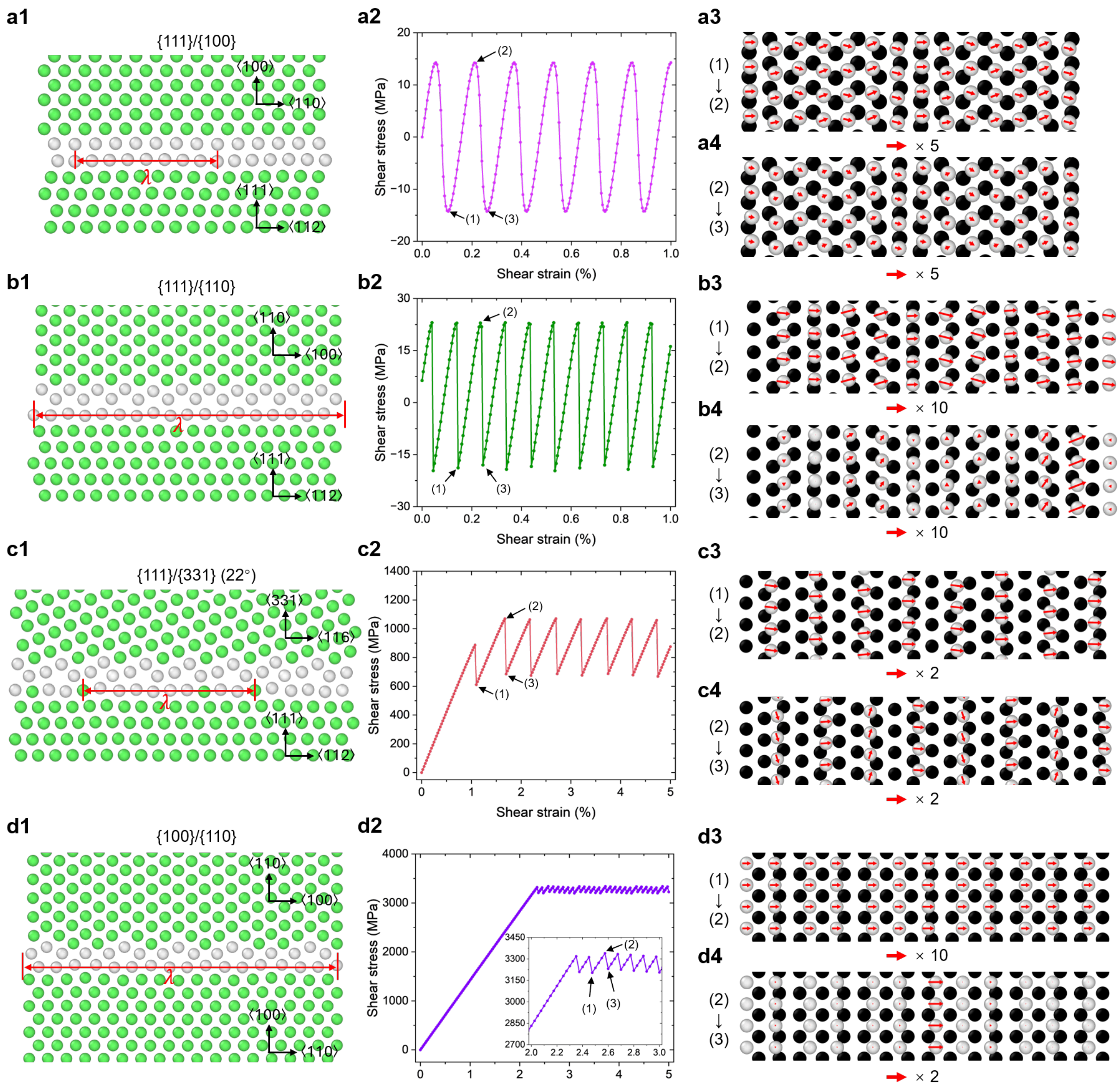

**Fig. 3 | Supercell atomic structures, shear stress-strain responses, and displacement fields of GBs exhibiting pure sliding**. **a1–a4**, {111}/{100} GB; **b1–b4**, {111}/{110} GB; **c1–c4**, {111}/{331}(22°) GB; **d1–d4**, {100}/{110} GB. For each boundary, panel 1 shows the relaxed atomic structure in a compact supercell with PBCs applied in the boundary plane, with crystallographic directions indicated in the two adjoining grains. Red arrows indicate the primary interfacial repeat length $\lambda$. Panel 2 shows the shear stress-strain curve, where states (1)-(3) define a representative sliding interval. Panels 3 and 4 present GB plan views of the boundary at states (2) and (3), respectively. Red arrows denote atomic displacement vectors from state (1) to (2) in panel 3 and from state (2) to (3) in panel 4. Vector amplification factors are specified alongside each configuration.

**Table 1 | Structural and deformation characteristics of four pure-sliding GBs.** Listed quantities include the SGBMD Burgers vector $\mathbf{b}_s$, the primary periodicity vector $\boldsymbol{\lambda}$ and its magnitude $|\boldsymbol{\lambda}|$, the average secondary periodicity $|\bar{\boldsymbol{\Lambda}}|$, and the GB shear strength without ($\tau_0$) and with ($\tau_s$) SGBMDs. Both $\mathbf{b}_s$ and $\boldsymbol{\lambda}$ are given in units of the lattice constant $a_0$, with the slip plane expressed in terms of the terminated plane of the reference grain. Where two shear strengths are listed, the upper and lower entries correspond to shear directions 1 and 2, respectively.

| GB configuration | $\mathbf{b}_s$ | $\boldsymbol{\lambda}$ | $\|\boldsymbol{\lambda}\|$ (nm) | $\|\bar{\boldsymbol{\Lambda}}\|$* (nm) | $\tau_0$ (MPa) | $\tau_s$ (MPa) |
|---|---|---|---|---|---|---|
| {111}/{100} | 1/14⟨112⟩{111}<br>1/8⟨110⟩{100} | 2⟨112⟩{111}<br>7/2⟨110⟩{100} | 1.73 | 5.97 | 14 | $3.5 \times 10^{-4}$ |
| {111}/{110} | 1/22⟨112⟩{111}<br>1/9⟨100⟩{110} | 9/2⟨112⟩{111}<br>11⟨100⟩{110} | 3.87 | 19.0 | 23 | $1.4 \times 10^{-2}$ |
| {111}/{331} (22°) | 1/8⟨112⟩{111}<br>1/20⟨116⟩{331} | 5/2⟨112⟩{111}<br>⟨116⟩{331} | 2.16 | 16.25 | 800<br>1,074 | -**<br>530 |
| {100}/{110} | 1/24⟨110⟩{100}<br>1/17⟨100⟩{110} | 17/2⟨110⟩{100}<br>12⟨100⟩{110} | 4.23 | 11.96 | 3,498 | 3,870 |

* $|\bar{\boldsymbol{\Lambda}}|$ represents the average spacing between neighboring SGBMDs. Individual spacings are integer multiples of $|\boldsymbol{\lambda}|$, typically $n|\boldsymbol{\lambda}|$ and $(n+1)|\boldsymbol{\lambda}|$, where $n$ is an integer[19]. The reported value is the average of these spacings weighted by their relative frequencies.

**The GB underwent successive structural transformations during shear loading, precluding determination of a well-defined shear strength.

Iterative workflow refinement enabled GBClaw to connect atomistic simulation data directly with mechanistic analysis of GB sliding. Here, we identified SGBMDs, previously recognized only conceptually[19], and quantified their influence on sliding resistance. Based on the results in Fig. 3, SGBMD Burgers vectors $\mathbf{b}_s$ were determined. Combining atomic displacement fields with structural recurrence analysis linked individual stress drops to discrete grain translations, allowing identification of the displacement-shift-complete (DSC) slip vector[44]. For the {111}/{331}(22°) GB (Fig. 3c), each stress drop was associated with a translation of 1/8⟨112⟩{111} relative to the {111} terminated grain. Four successive drops accumulated a slip of 1/2⟨112⟩{111}, carrying each atom on the {331} terminating plane to a crystallographically equivalent site along the sliding direction. The translation associated with a single stress drop was the shortest displacement that restored the interfacial structure and was therefore identified as the SGBMD Burgers vector $\mathbf{b}_s$.

Applying the same analyses to all four GBs yielded different $\mathbf{b}_s$ assignments (Table 1), reflecting their distinct bicrystallographies.

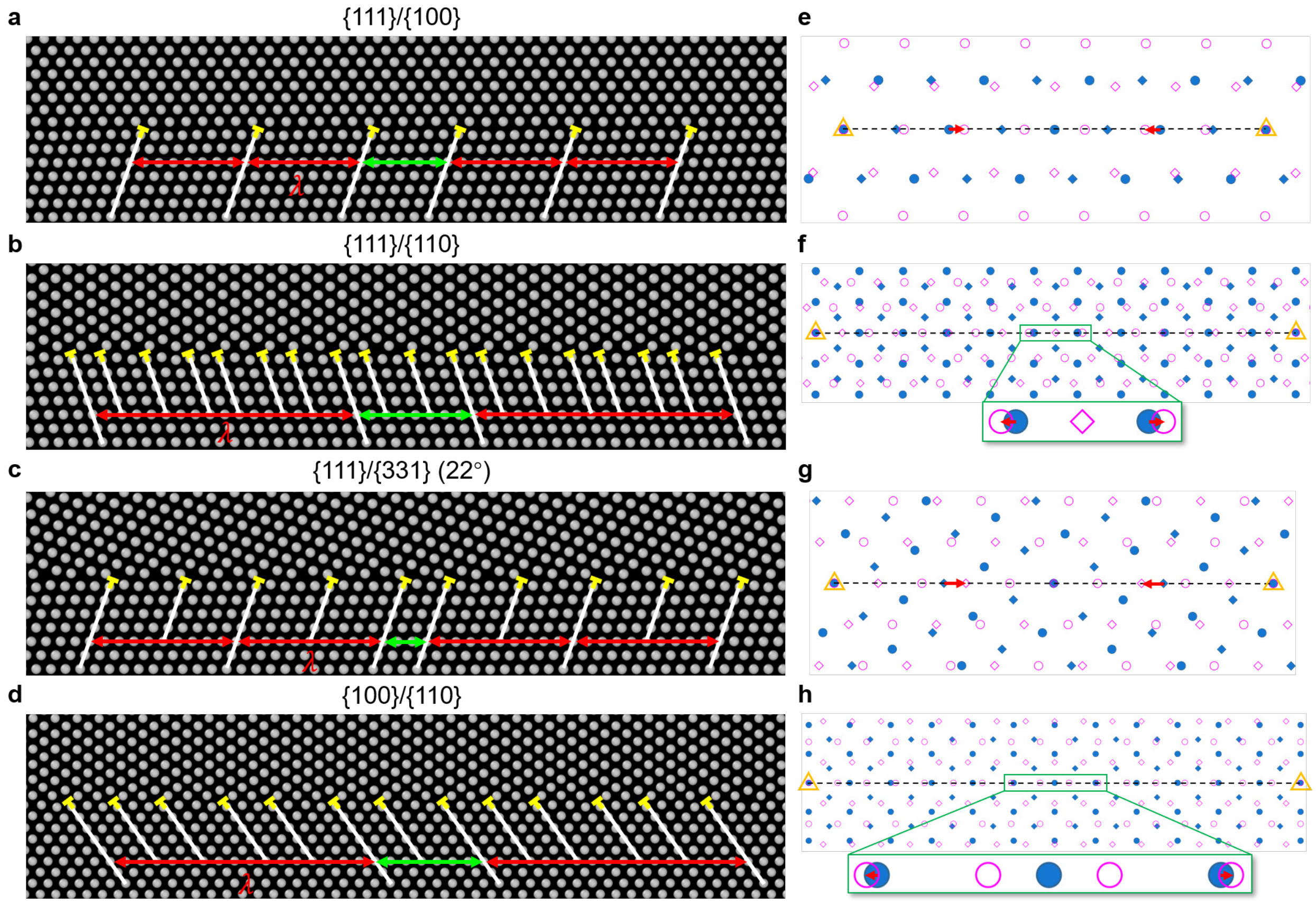

**Fig. 4 | Identification of PGBMDs and SGBMDs and determination of SGBMD Burgers vectors in four GBs undergoing pure sliding. a-d**, Geometric analysis of projected atomic-column arrangements in simulated atomic configurations of four GBs: **a**, {111}/{100}; **b**, {111}/{110}; **c**, {111}/{331}(22°); **d**, {100}/{110}. White lines trace extra {111} half-planes. Yellow "⊥" symbols denote PGBMD cores. Red double-headed arrows indicate the primary structural period λ. Green double-headed arrows indicate the SGBMD core span. **e-h**, Corresponding near-CSL dichromatic patterns for the GBs in **a-d**. Red arrows indicate the SGBMD Burgers vectors $\mathbf{b}_s$ for the two opposite sliding directions.

Using extended bicrystal models of the four GBs, a geometric method was established to identify PGBMDs and SGBMDs in GB structures and calculate their Burgers vectors. Atomic configurations of the simulated bicrystals were projected onto a 2D plane to enable comparison with HRTEM images. As shown in Fig. 4a-d, PGBMD cores were identified at the terminations of extra {111} half-planes, marked by white lines; these cores were denoted by yellow "⊥" symbols. The shortest complete repeat of the PGBMD arrangement defined the local GB structural period λ indicated by red double-headed arrows. Localized interruptions in this periodic arrangement indicated SGBMD cores, with their spans marked by green double-headed arrows. Each SGBMD Burgers vector was calculated from the residual mismatch across its core span relative to the periodic PGBMD arrangement in a neighboring reference span of length λ. Volumetric-strain maps provide a complementary check of the simulated defect assignments (Supplementary Fig. S5 and Supplementary Note 2). Researcher guidance subsequently prompted GBClaw to automate SGBMD identification and Burgers vector determination from projected atomic-column arrangements in individual HRTEM images. Application to *in situ* images of a {111}/{331}(22°) GB during sliding revealed agreement between the experimental atomic-column arrangements and the simulated defect configurations, supporting the geometric defect assignment (Fig. 4a-d, Supplementary Note 2, Figs. S6 and Table S3).

Further exchanges led to an independent crystallographic analysis of SGBMD Burgers vectors using near-CSL dichromatic patterns (Supplementary Note 3 and Fig. S7). For each bicrystal, the two grain lattices were superimposed using the primary structural period determined from atomistic simulation. Biaxial elastic strains and a relative lattice rotation were solved for producing a common periodic cell with an exact CSL. This construction provided a commensurate approximation to the original incommensurate bicrystal. As shown in Fig. 4e-h, the shortest nonzero vector within each GB plane connecting sites in the two grains at the same depth along the projection direction was identified as the minimum DSC translation. These vectors, indicated by red arrows, agreed with the corresponding Burgers vectors $\mathbf{b}_s$. Agreement among structural recurrence, geometric defect analysis, and the near-CSL construction corroborated the SGBMD identifications and Burgers vector assignments. Together, these analyses established a workflow for interpreting both atomistic configurations and experimental HRTEM images, illustrating how researcher guidance enabled GBClaw to conduct and cross-check the mechanistic analysis of SGBMDs.

Within these general asymmetric tilt GBs, PGBMDs are typically sessile in the boundary plane, acting as structural obstacles to the motion of other defects. In contrast, SGBMDs serve as mobile carriers of interfacial sliding. To examine how SGBMDs affect GB sliding, the structure and LAMMPS agents evaluated the shear strengths of bicrystal models containing arrays of these defects. Each supercell incorporated multiple SGBMDs whose Burgers vectors summed to a lattice translation along the boundary, preserving crystallographically equivalent interfacial registry across periodic images. Under PBCs, these models represent GBs containing regular SGBMD arrays. For the {111}/{331}(22°) GB, sliding proceeded through the intermittent glide of individual dislocations or localized sub-arrays (Supplementary Note 4 and Fig. S8). The presence of SGBMDs substantially reduced the shear strength from $\tau_0 = 1{,}074$ MPa in the SGBMD-free compact model to $\tau_\mathrm{s} = 530$ MPa in the model with SGBMDs. Table 1 compares $\tau_0$ and $\tau_\mathrm{s}$ for all four GBs alongside the corresponding magnitudes of the secondary periodicity vector $\mathbf{\Lambda}$. These results demonstrate that the presence of SGBMDs changes the GB shear strength, highlighting their role in GB sliding resistance and the importance of accounting for defect hierarchies in atomistic models. Further analysis of how spatially non-uniform distributions of SGBMDs influence sliding in general GBs will be reported separately.

Dynamic replanning enabled GBClaw to leverage researcher guidance to provide mechanistic insights into shear asymmetry and deformation-mode selection. The analysis agent compared the 22° and 49° misorientation variants of the {111}/{331} GB. The two variants arise from two {111} planes whose inclination angles differ by 71° about the ⟨110⟩ tilt axis. Both variants share a primary periodicity $\boldsymbol{\lambda} = 5/2\langle 112\rangle\{111\}$ and an SGBMD Burgers vector $\mathbf{b}_\mathrm{s}$ = 1/8⟨112⟩{111} (Table 1, Table S4, and Supplementary Fig. S9). When sheared along direction 1 (i.e., the mode of lower shear strength), the 49° variant exhibited a pure-sliding response similar to that of the 22° variant. Its repeated stress drops produced a comparable sawtooth profile, albeit at different stress levels (Figs. 3c2 and S10a). When shear was applied along direction 2, however, the 49° GB underwent shear-coupled migration (Fig. S10b), whereas the 22° GB maintained pure sliding. Dichromatic mapping based on near-CSL constructions revealed different lattice registries for the two variants (Supplementary Figs. S7c and S9a), suggesting that differences in subsurface stacking sequences and second-nearest-neighbor coordination likely underlie the distinct shear pathways (Supplementary Note 5 and Figs. S9, S10). This comparison further illustrates how GBClaw bridges crystallographic characterization with the local atomic mechanisms governing shear responses. In addition, dynamic replanning enabled GBClaw to leverage researcher guidance to investigate an anomalous shear response and attribute the initially high sliding resistance to a metastable local structure formed during bicrystal construction and relaxation (Supplementary Note 6 and Fig. S11). This investigation demonstrates how human-in-the-loop replanning enables GBClaw to unravel unexpected simulation phenomena and pinpoint their atomistic reconstruction mechanisms.

## Discussion

GBClaw served as an effective co-scientist, enabling LLM-driven workflows for atomistic GB research and accelerating mechanistic discovery that would previously have been difficult to achieve within practical timeframes. These workflows resolved misfit-dislocation hierarchies, characterized distinct deformation modes, and quantified substantial variations in athermal shear strengths across asymmetric incommensurate GBs. For SGBMDs, which had previously been recognized only conceptually[19], researcher-guided workflow refinement established a general approach for identifying these interfacial defects and determining their Burgers vectors from simulated atomic configurations and experimental HRTEM images. The assignments were corroborated by complementary analyses of structural recurrence associated with stress drops during simulated sliding and of DSC translations in near-CSL reference structures. Further simulations revealed significant differences in shear strengths between GB models with and without SGBMDs.

Within general asymmetric tilt GBs, PGBMDs are typically sessile in the boundary plane, acting as structural obstacles to motion of other defects. In contrast, SGBMDs serve as mobile carriers controlling interfacial sliding. These findings underscore the critical need to represent the relevant defect hierarchies when using high-fidelity first-principles calculations or machine-learned potentials to design GB composition and structure for enhanced mechanical performance. Improving the description of interatomic interactions alone cannot compensate for omitting structural defects that govern deformation. Furthermore, dynamic replanning directed follow-up calculations that resolved the structural origins of the shear asymmetry in the {111}/{331}(49°) GB and the anomalous shear response in the {111}/{110} GB. Together, these examples illustrate how automation scales systematic exploration while researcher guidance ensures physical validation and mechanistic discovery.

Several methodological constraints of the underlying atomistic simulations warrant consideration. The current implementation employed compact supercells with PBCs within the GB plane to evaluate shear strengths with and without SGBMDs. These spatial constraints restrict the spatial distribution and long-range interactions of line defects, potentially suppressing deformation pathways active in extended boundaries. The calculated strengths therefore represent athermal instability thresholds for the modeled boundary configurations, rather than universal upper bounds. Likewise, while NEB calculations delineate minimum energy paths between selected states, they do not explicitly account for finite-temperature dynamics. Addressing these limitations will require large-scale molecular dynamics simulations to capture interactions among multiple defects, alongside enhanced sampling or kinetic rate-theory methods to evaluate thermally activated processes.

GBClaw's modular architecture and hierarchical memory provide an extensible foundation for addressing these challenges. Specialized agents currently under development focus on the adaptive refinement of machine-learned potentials tailored to complex GB configurations and on experimental validation through atomic-resolution characterization. Integrating these agents through standardized data interfaces will enable GBClaw to coordinate high-throughput structural screening, potential refinement, atomistic simulation, and experimental validation within a single closed-loop system.

To balance flexibility and reliability, GBClaw combines standardized scientific tools with custom code dynamically generated by its coding agent. Predefined tools establish reproducible workflows for routine operations[45], whereas custom code provides the adaptability needed for specialized analyses and error recovery[16]. However, token consumption aside, repeated code generation risks introducing algorithmic inconsistencies across runs[46]. Converting validated scripts into persistent, structured tools after their physical assumptions, inputs, and numerical behavior have been checked mitigates this variability. As highlighted by the domain-knowledge pitfalls reported for MatClaw[6], successful execution does not guarantee physical correctness. Multi-layer verification, physical sanity checks, and human expert oversight remain indispensable.

In summary, GBClaw demonstrates that agentic AI can effectively connect high-throughput simulations with deep mechanistic discovery, while maintaining full provenance across models, execution records, and analytical outputs. By elucidating the role of misfit-dislocation hierarchies in interfacial deformation, this work advances a unified, defect-based framework for investigating complex interfaces. Beyond homophase boundaries, GBClaw provides a generalizable framework for investigating heterophase interfaces, van der Waals heterostructures, and interfacial friction in incommensurate solid-state contacts.

## Methods

### Implementation of GBClaw

GBClaw was implemented in Python using the open-source LangChain and LangGraph

frameworks[17] for agent coordination and DeepAgents for task-specific code execution. The system used both GPT-5.6 Sol and GPT-5.6 Luna, with task-dependent model routing selecting the model according to the complexity and reasoning requirements of each task. Specialized agents accessed LAMMPS[26], OVITO[27], and ASE[28] through Python-based tool interfaces. Computational results were linked to persistent research records and supporting artifacts. Supplementary Table S5 lists the tools available to each agent, and Supplementary Fig. S4 illustrates a representative researcher-guided investigation. Further implementation details and software dependencies are documented in the source code repository identified in the Code Availability statement.

**GB selection and model construction**

General asymmetric ⟨110⟩ tilt GBs were selected near the exact CSL reference boundaries with $\Sigma < 50$ listed in Supplementary Table S2. Crystallographically equivalent plane pairs were grouped according to cubic symmetry, and one representative orientation was retained from each group. The {111}/{331}(49°) and {111}/{221}(55°) variants were added to the 12 selected plane-family pairs, yielding 14 GB configurations. Ni bicrystals were constructed with the tilt axis along the out-of-plane direction. Interatomic interactions were described using the angular-dependent Ni-Cr potential of Howells and Mishin[42]. The lattice parameter of Ni $a_0$ was 3.52 Å. Compact periodic supercells were selected with an integer multiple of the primary periodicity along the horizontal direction, as illustrated in Fig. 2c, d. These supercells retained the local interfacial motif while excluding SGBMDs. Residual mismatch was accommodated by elastic strain, with both the atomic coordinates and cell lengths allowed to relax. Supplementary Table S4 reports the crystallographic orientations, lattice repeat counts, relaxed supercell dimensions and residual lattice mismatch.

Each compact supercell contained one GB between upper and lower free surfaces. PBCs were applied within the GB plane, and vacuum regions with an initial thickness of 20 Å were introduced above and below the free surfaces. Before relaxation, overlapping atoms separated by less than $0.3a_0$ across the GB were removed. Structural relaxation was performed in LAMMPS[26] through four cycles, each alternating between atomic relaxation at fixed supercell dimensions and cell-length relaxation toward zero normal stress. Conjugate gradient energy minimization used a force tolerance of $10^{-6}$ eVÅ$^{-1}$, with maximum limits of 1,000 iterations and 10,000 force evaluations for each minimization. The energy-based convergence criterion was disabled.

**Shear simulations**

Shear displacement was applied incrementally along the horizontal direction. Boundary layers with a thickness of 3 Å were defined at the upper and lower surfaces of the bicrystal, and their atomic membership was retained throughout loading. At each load increment, the upper layer was displaced horizontally while the lower layer remained fixed in that direction. During energy minimization, the in-plane coordinates of both boundary layers were constrained, while their coordinates normal to the GB plane were allowed to relax. The cell dimensions remained fixed throughout loading. The shear strain was defined as $\gamma = \Delta u/H_0$, where $\Delta u$ is the relative horizontal displacement between the centers of the boundary layers, and $H_0$ is their initial separation normal to the GB plane.

The default forward displacement used a strain increment of $\Delta\gamma = 2.5\times10^{-4}$ over 200 increments. Each increment was followed by conjugate gradient energy minimization with a force tolerance of $10^{-6}$ eVÅ$^{-1}$. The energy-based convergence criterion was disabled, and each minimization was limited to 3,000 iterations and 10,000 force evaluations. Shear stress was calculated as the negative of the LAMMPS pressure-tensor component $P_{xy}$ and rescaled by the ratio of the initial supercell volume to the initial atomic bounding-box volume, thereby excluding the vacuum contribution. The shear strength in Fig. 2e was defined by the peak stress within the subsequent steady response after excluding the initial loading transient. Directions 1 and 2 denote the directions with lower and higher shear resistance, respectively. For the {111}/{100} and {111}/{110} GBs shown in Fig. 3a, b, the strain increment was reduced fivefold to $5\times10^{-5}$ to resolve the sliding response more finely. Simulations with periodic arrays of SGBMDs are described in Supplementary Note 4 and Supplementary Fig. S8.

**Loading-unloading cycle and NEB calculations**

The {111}/{110} GB was loaded beyond the initial stress drop and unloaded to zero applied shear strain using the same displacement-control procedure. The loading range and number of increments are provided in Supplementary Note 6. The relaxed initial and unloaded configurations were used as endpoints for a strain-controlled NEB calculation[23,24] at zero applied shear strain. Atoms were matched by their identifiers, and 144 replicas, including both endpoints, were initialized through linear interpolation using periodic minimum-image displacements. The minimum energy path was obtained using the FIRE algorithm with a spring coefficient of 1.0 eVÅ$^{-2}$ and a minimization timestep of 0.01 ps. The conventional NEB and climbing-image stages were each limited to 10,000 iterations and used a force tolerance of $10^{-6}$ eVÅ$^{-1}$. The energy-based convergence criterion was disabled. Energies were reported relative to the relaxed initial endpoint. Supplementary Table S4 provides supercell dimensions.

**Code availability**

The GBClaw source code is provided under the MIT License on GitHub (https://github.com/ZhuGroup-GT/GBClaw). The recorded demo project can be visited at https://zhugroup.gatech.edu/gbclaw-demo.

**Data availability**

All relevant data are included within the manuscript, Supplementary Information, and GitHub repository. The detailed simulation artifacts are available from the corresponding author upon request.

**Author contributions**

YL, KD and TZ conceived research. YL, KD, XG, SO and TZ conducted research. YL wrote the initial manuscript. WC and TZ revised the manuscript. All the authors contributed to the discussion of the results.

**Competing interests**

The authors declare no competing interests.

**Acknowledgements**

YL, KD, YY, JK, OP and TZ acknowledge support by the U.S. Department of Energy (DOE), Office of Science, Basic Energy Sciences (BES) Materials Science and Engineering (MSE) Division under Award #DE-SC0018960. WC acknowledges support by the National Science Foundation (Award numbers 2516550 and 2331132). XG and SO acknowledge support by Japan Society for the Promotion of Science (JSPS) KAKENHI Grant No. JP23K20037. SO acknowledges support by Japan Science and Technology Agency (JST) ARiSE Grant Numbers JPMJAR2601 and JPMJAR2603 and Ministry of Education, Culture, Sport, Science and Technology of Japan (MEXT) Program Grant Number JPMXP1122684766.

# Supplementary Information

# Mechanistic discovery of grain boundary structure and deformation driven by agentic AI

Yazhuo Liu[a,†], Kunqing Ding[a,†], Xiaoguo Gong[b], Zifeng Wang[a], Yichen Yang[a], Yifeng Che[a], Wen Chen[d,e], Josh Kacher[c], Olivier Pierron[a], Shigenobu Ogata[b], Ting Zhu[a,c,]*

[a]Woodruff School of Mechanical Engineering, Georgia Institute of Technology, Atlanta, GA, 30332, USA

[b]Department of Mechanical Science and Bioengineering, Graduate School of Engineering Science, The University of Osaka, Osaka, 560-8531, Japan

[c]School of Materials Science and Engineering, Georgia Institute of Technology, Atlanta, GA, 30332, USA

[d]Department of Aerospace and Mechanical Engineering, University of Southern California, Los Angeles, CA, USA

[e]Mork Family Department of Chemical Engineering and Materials Science, University of Southern California, Los Angeles, CA, USA

† These authors contributed equally to this work.

* Corresponding author: ting.zhu@me.gatech.edu

## Supplementary figures

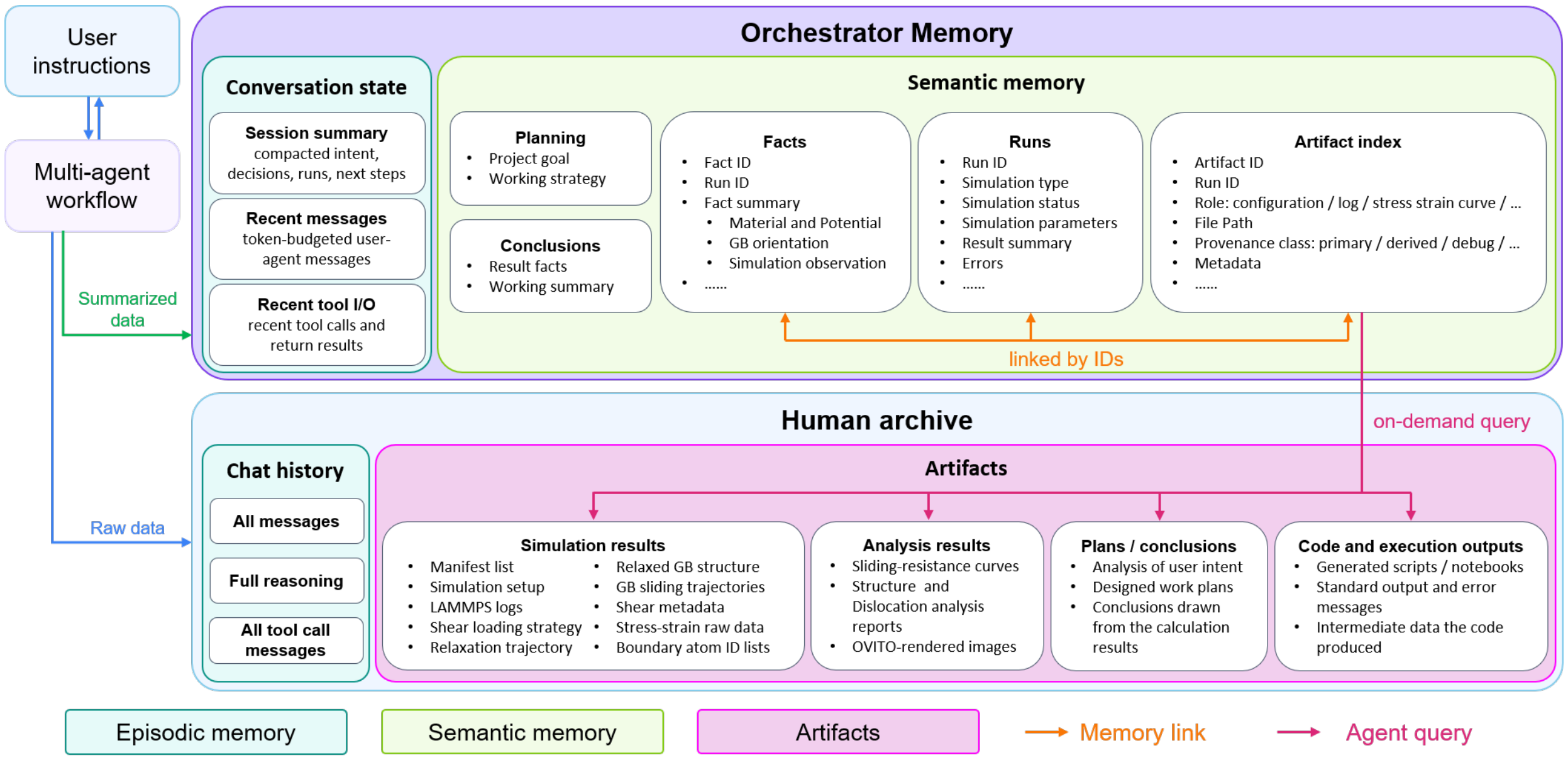

**Fig. S1 | Memory organization and identifier-based information flow in GBClaw.** Conversation state retains the session summary, recent messages, and recent tool inputs and outputs. Semantic memory stores project plans, conclusions, facts, run records and an artifact index. Shared run IDs connect fact records with simulation conditions and outputs, while artifact IDs link indexed entries to the corresponding files for retrieval on demand. Complete interaction and tool execution histories are archived for human review. Green and blue arrows represent summarized and raw information, respectively. Orange arrows denote memory links, and magenta arrows indicate artifact queries.

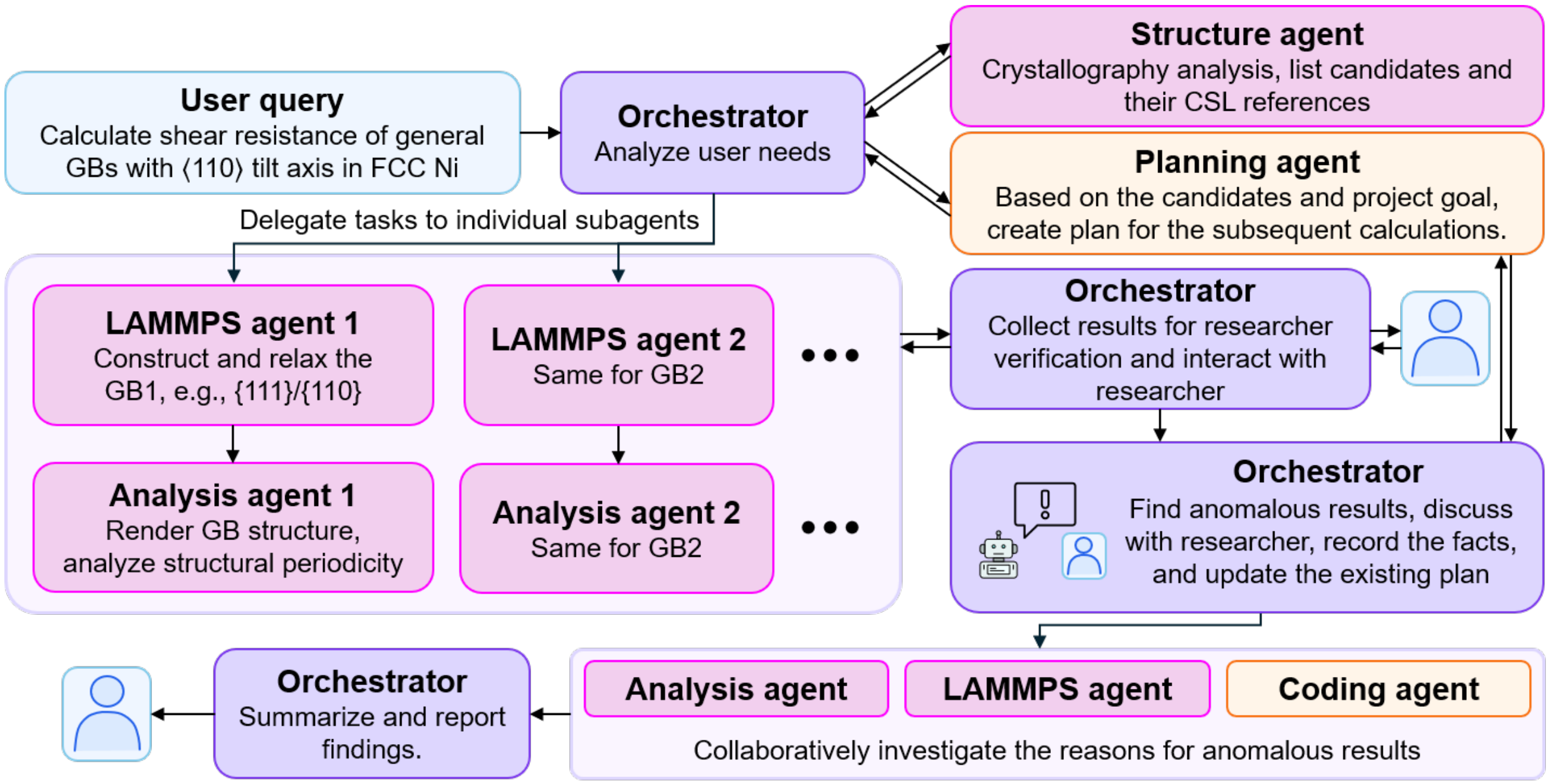

**Fig. S2 | Agentic GBClaw workflow for systematic screening and mechanistic analysis of GB shear strength.** The workflow combines automated exploration with human-in-the-loop domain guidance across six key stages: crystallographic screening, simulation planning, parallel bicrystal construction and structural analysis, researcher review, dynamic replanning, and targeted investigation of anomalous shear responses in general asymmetric ⟨110⟩ tilt GBs.

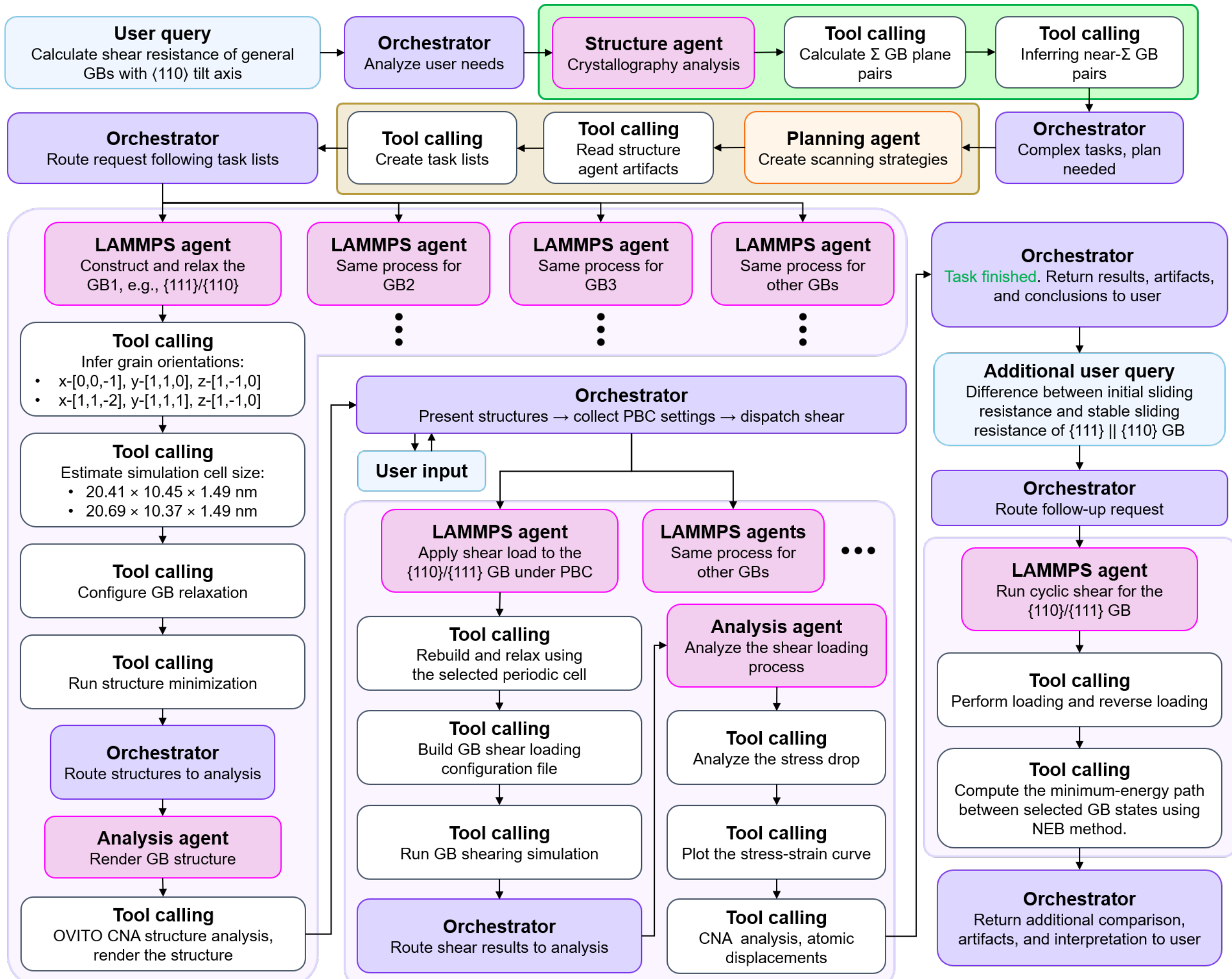

**Fig. S3 | Detailed GBClaw workflow for investigating GB shear strength.** Automated tasks and researcher guidance connect crystallographic screening, simulation planning, parallel bicrystal construction and relaxation, selection of periodic boundary conditions, shear simulations, and structural and mechanical analysis of general asymmetric ⟨110⟩ tilt GBs. The {110}/{111} GB illustrates an adaptive investigation, in which an unexpected shear response prompted researcher-directed follow-up analyses using cyclic shear simulations and NEB calculations.

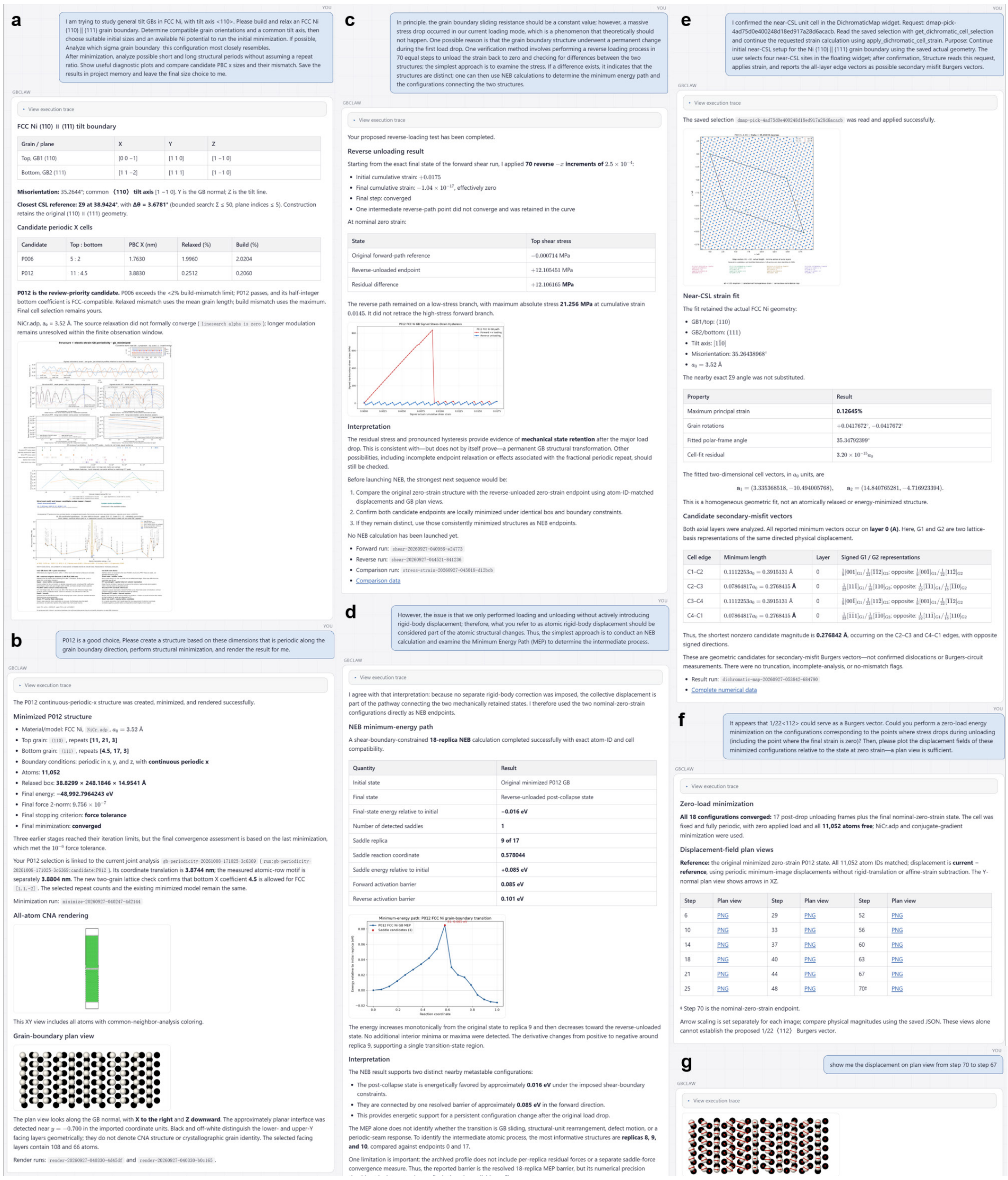

**Fig. S4 | Representative researcher-GBClaw interaction underlying the workflow in Fig. S3.** **a**, Selection of bicrystal orientations and periodic-cell candidates. **b**, Construction, relaxation and visualization of the selected bicrystal. **c**, Unloading to zero applied strain and comparison with the initial state. **d**, NEB calculation between the initial and unloaded configurations. **e**, Near-CSL dichromatic analysis of SGBMD Burgers vectors. **f**, Relaxation of selected configurations and comparison of their displacement fields. **g**, Atomic displacements during a selected unloading interval. Detailed tool calls, reasoning summaries and computational outputs are available in the accompanying online demonstration (https://zhugroup.gatech.edu/gbclaw-demo). The displayed NEB setup is illustrative. Final NEB settings are given in Methods and Supplementary Note 6.

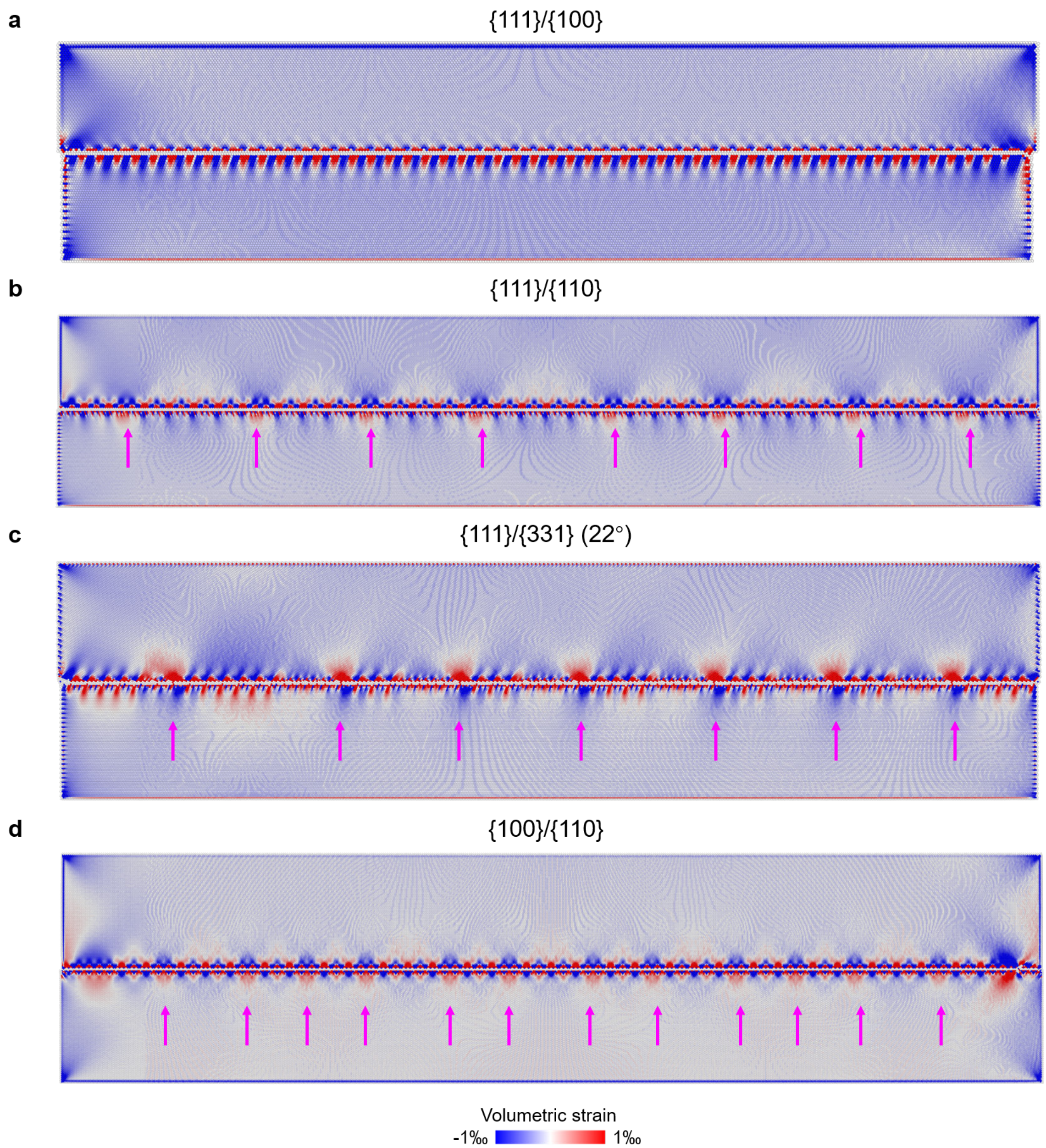

**Fig. S5 | Atomic volumetric strain in extended bicrystals containing primary and secondary GB misfit dislocations.** **a**, {111}/{100} GB. **b**, {111}/{110} GB. **c**, {111}/{331}(22°) GB. **d**, {100}/{110} GB. All configurations are viewed along the common ⟨110⟩ tilt axis. Atoms are colored by volumetric strain. Arrows in **b** to **d** indicate localized SGBMD cores. The strain map in **a** does not resolve individual SGBMD cores. This lack of strain localization is due to core spreading, consistent with the low sliding resistance.

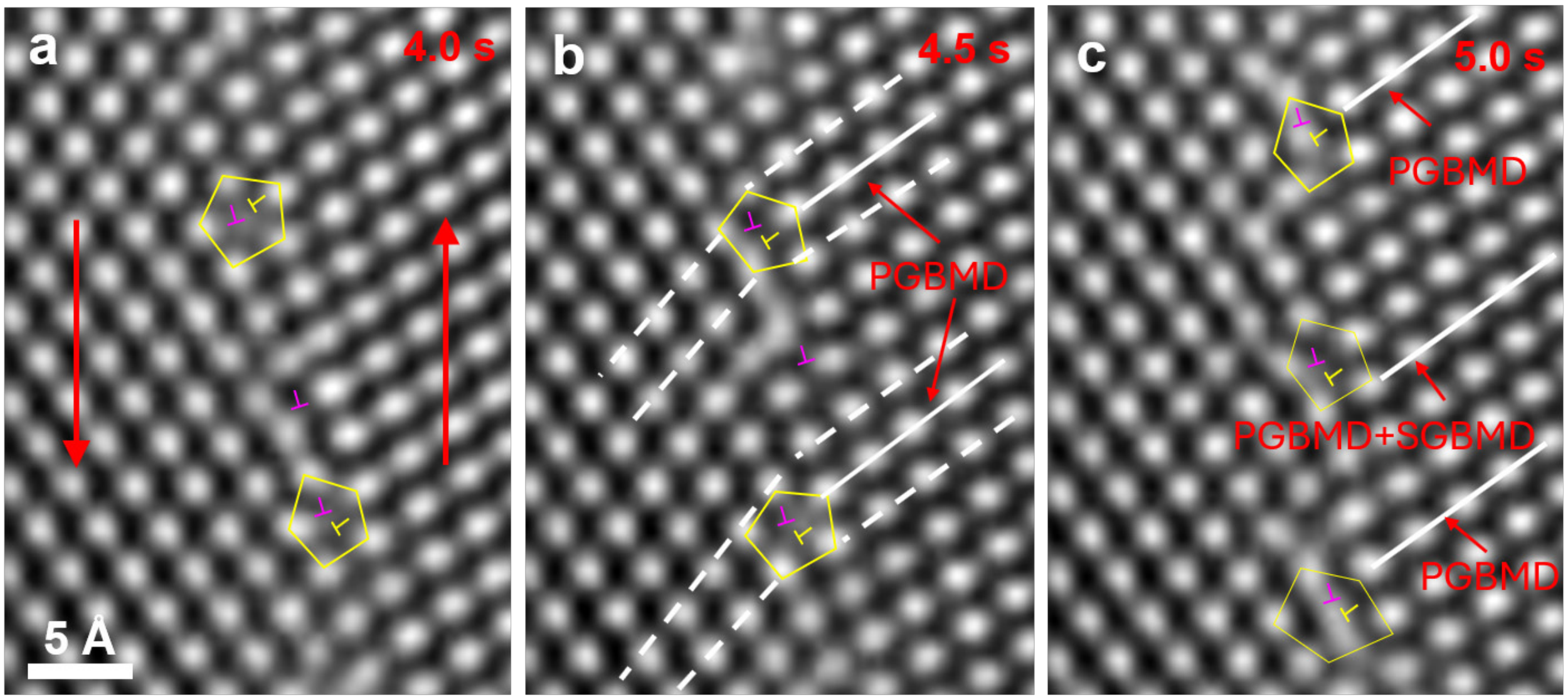

**Fig. S6 | Identification of SGBMDs from projected atomic-column arrangements in *in situ* HRTEM images of a {111}/{331}(22°) GB undergoing sliding**, adapted from Wang et al[36]. **a, b**, Each image shows two PGBMDs. In **a**, arrows indicate the sliding direction. In **b**, solid white lines trace extra {111} half-planes, whose interfacial terminations identify the PGBMD cores. **c**, An SGBMD identified by an additional {111} half-plane. The associated atomic-column arrangement is consistent with the simulated structure shown in Fig. 4c.

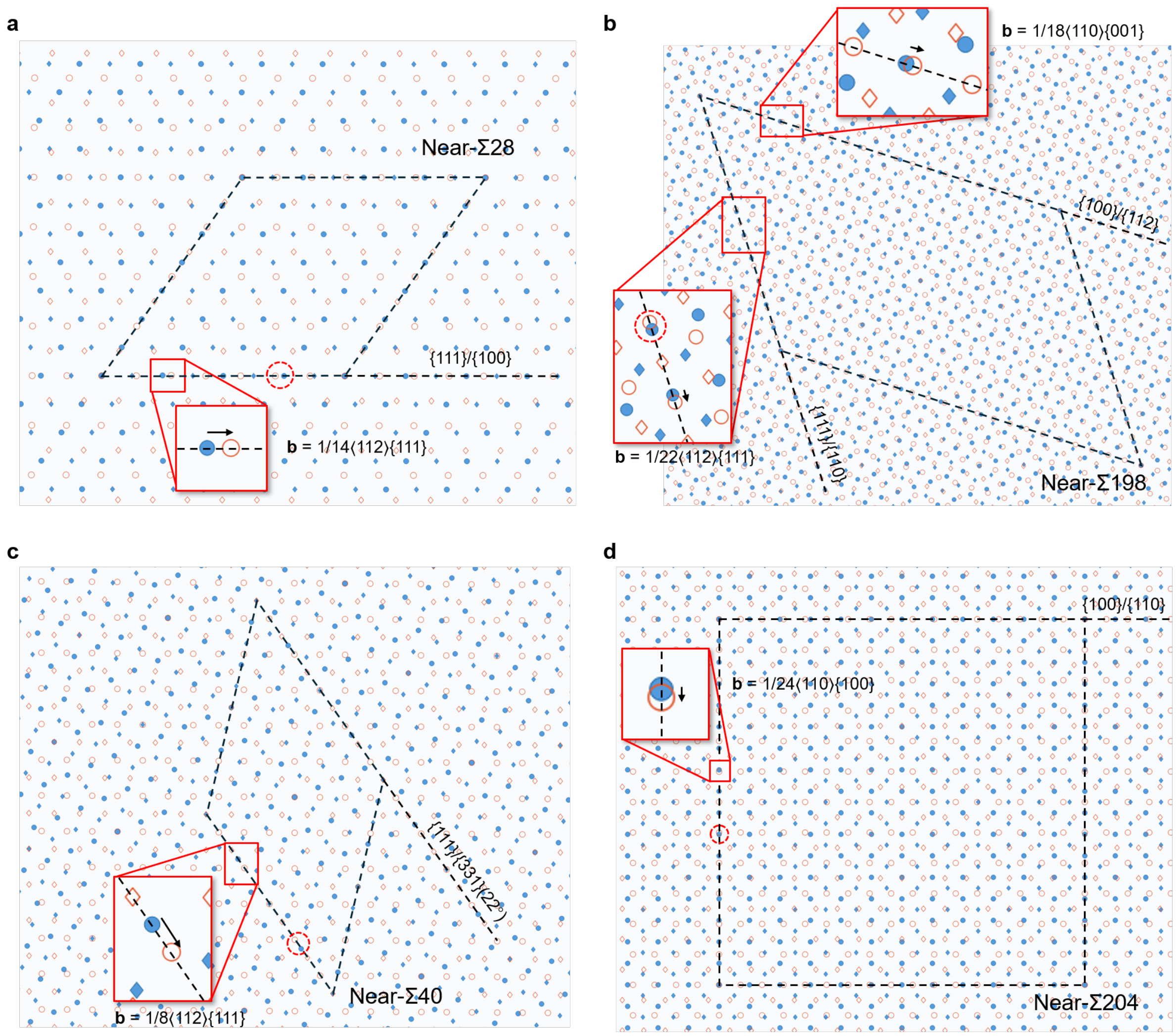

**Fig. S7 | Determination of SGBMD character from near-CSL dichromatic patterns of the four pure-sliding GBs.** **a**, {111}/{100} GB; **b**, {111}/{110} GB; **c**, {111}/{331}(22°) GB; **d**, {100}/{110} GB. Blue and orange symbols distinguish the two grain lattices, while circles and diamonds distinguish alternating atomic layers along the ⟨110⟩ tilt axis. Black dashed lines indicate the near-CSL cells and GBs. Black arrows indicate the shortest DSC translations connecting lattice sites in the two grains within the same atomic layer. These translations define the SGBMD Burgers vectors and agree with those obtained from the projected atomic-column analysis in Fig. 4a-d and Table S3. Red dashed circles mark symmetry-related site pairs that produce DSC translations equal in magnitude but opposite in sign to those indicated in red solid boxes.

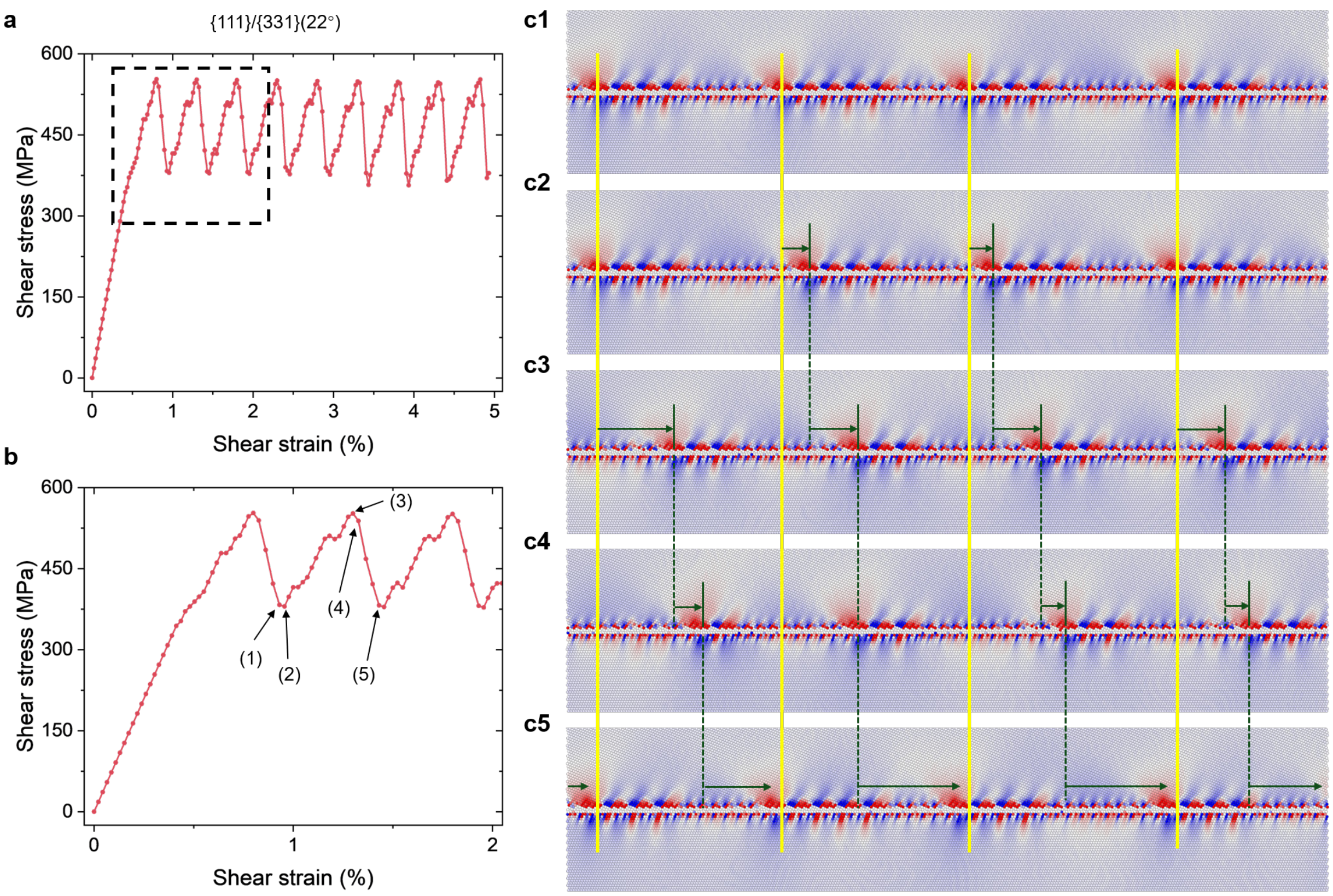

**Fig. S8 | Boundary sliding through nonuniform glide of SGBMDs in the {111}/{331}(22°) GB.** **a**, Shear stress as a function of applied shear strain. **b**, Magnified view of the stress response, with states (1) to (5) marking a representative sliding cycle. **c1** to **c5**, Atomic configurations corresponding to the marked states. Large red regions identify SGBMD cores. Yellow lines mark equivalent SGBMD positions at the beginning and end of the cycle. Green lines track dislocation positions, and arrows indicate movement between consecutive configurations.

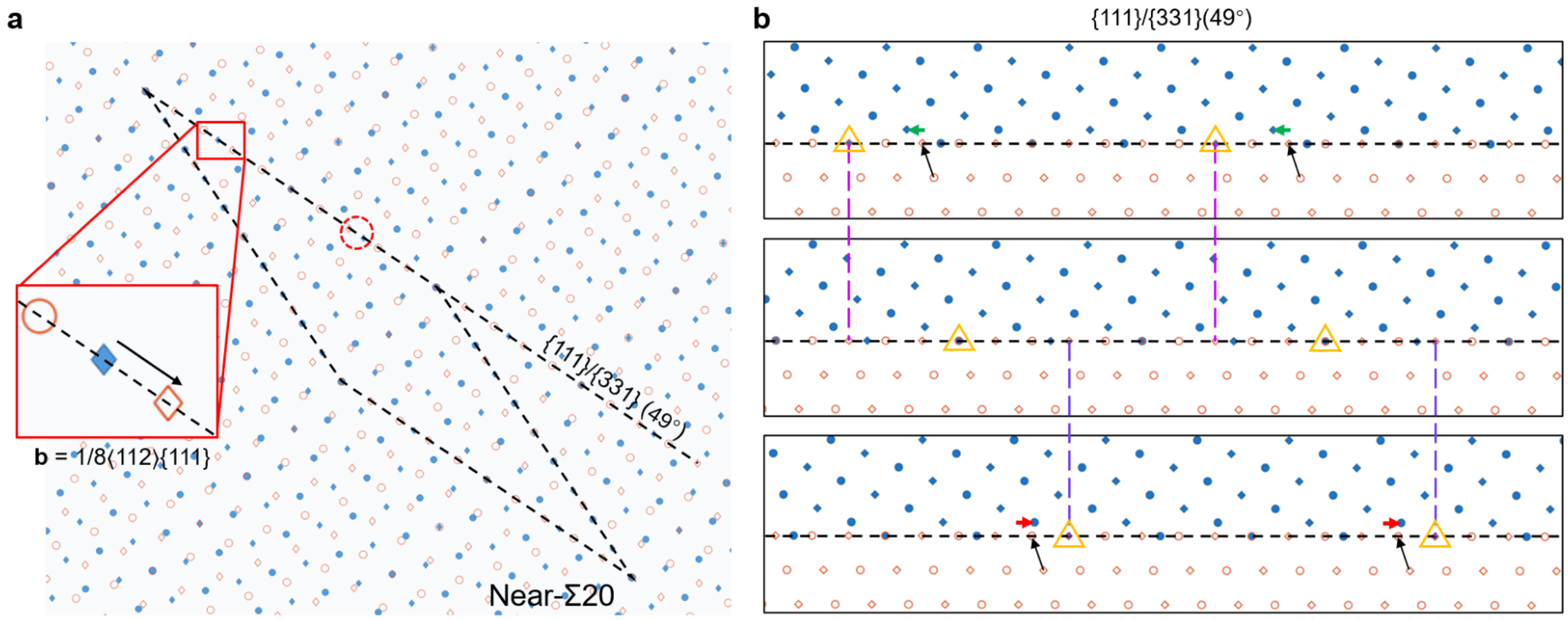

**Fig. S9 | SGBMD Burgers vector derived from the near-CSL pattern of the {111}/{331}(49°) GB. a,** Near-CSL dichromatic pattern. Blue and orange symbols distinguish between the two grain lattices, while circles and diamonds distinguish atomic layers along the ⟨110⟩ tilt axis. Dashed outlines indicate the near-CSL cell. The black arrow in the inset marks the shortest DSC vector connecting sites of the two lattices, yielding $\mathbf{b_s}$ = 1/8⟨112⟩{111}, identical to that of the 22° boundary in Fig. S7c. The dashed circle marks a symmetry-related pair of sites with opposite relative offsets in the reference pattern. **b**, Local interfacial configurations, with the blue and orange grains positioned above and below the GB, respectively. Taking the middle configuration as the reference, the orange atoms are held fixed, while the blue atoms are translated left by -$\mathbf{b}_s$ to obtain the upper configuration and right by +$\mathbf{b}_s$ to obtain the lower configuration. Green and red arrows denote these opposite tangential translations, each equal in magnitude to the Burgers vector of the SGBMD ($\mathbf{b}_s$). Black arrows indicate the resistance from second-nearest-neighbor atoms beneath the terminating plane. Yellow triangles highlight coincidence lattice sites, and vertical dashed guides provide positional references across the three registries.

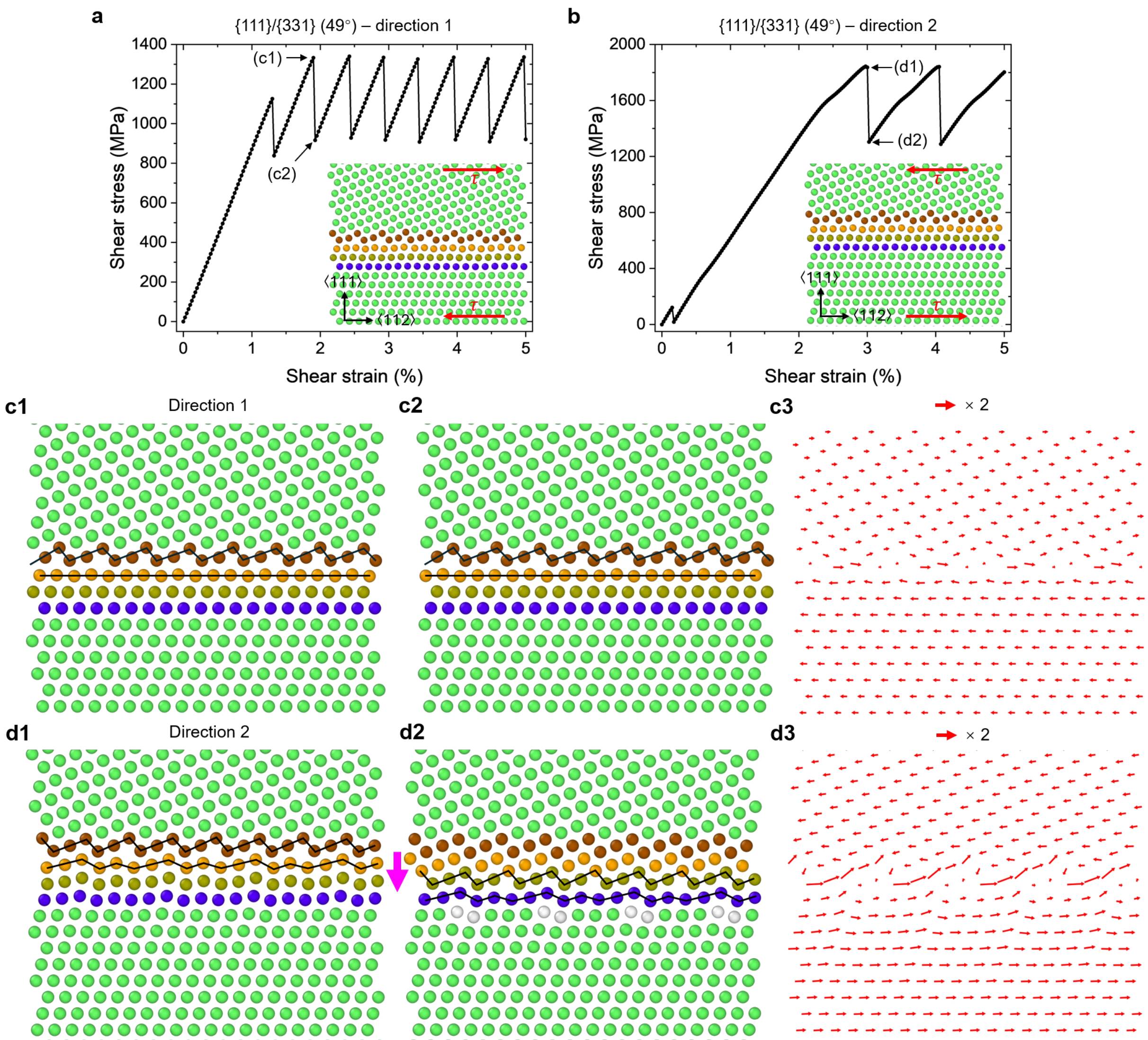

**Fig. S10 | Shear asymmetry in the {111}/{331}(49°) GB. a, b**, Shear stress as a function of applied shear strain along directions 1 and 2, respectively. Insets show the initial boundary structures and shear directions. **c1-c3,** Pure GB sliding when sheared along direction 1. **d1-d3,** Shear-coupled GB migration when sheared along direction 2. **c1, d1** and **c2, d2** show configurations before and after, respectively, the stress drops indicated in **a** and **b**. **c3, d3,** Corresponding atomic displacement vectors, amplified twofold for clarity. Selected atomic rows are colored to highlight structural changes. Black lines mark terminating planes. The magenta arrow indicates downward boundary migration during shear along direction 2. All configurations are viewed along the ⟨110⟩ tilt axis.

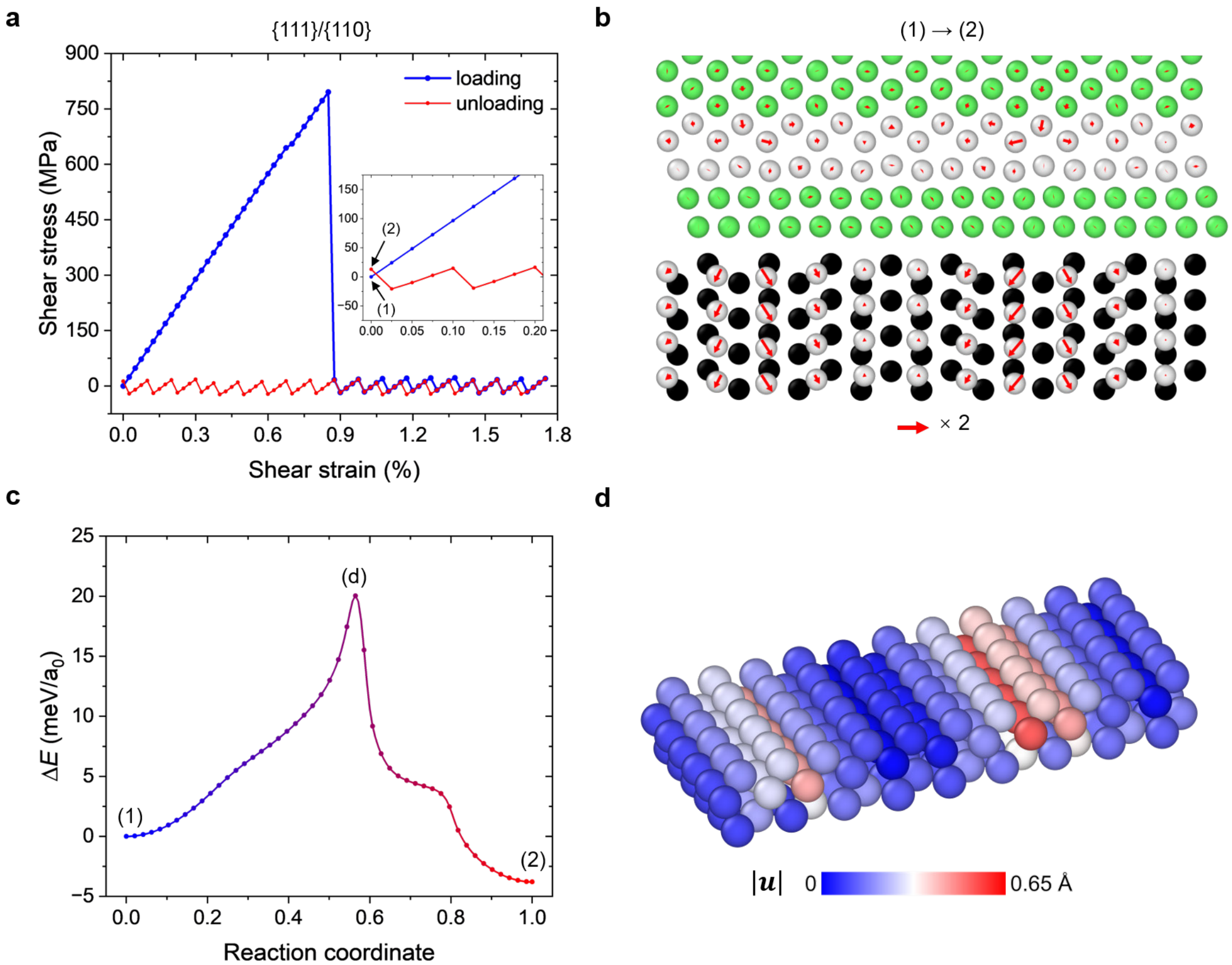

**Fig. S11 | Loading-induced local reconstruction and associated energy barrier in the {111}/{110} GB. a.** Shear stress-strain response during loading (blue) and unloading (red). The inset shows the residual shear stress of the initial relaxed configuration (1) and the final unloaded configuration (2), both evaluated at zero applied shear strain. **b.** Edge-on (upper) and plan (lower) views of configuration (2). Red arrows indicate atomic displacement vectors from configuration (1) to (2), scaled by a factor of two. **c.** Minimum energy path from an NEB calculation connecting configurations (1) and (2) at zero applied shear strain. **d.** Atomic configuration at the saddle point in **c**. Atoms are colored by their displacement magnitude relative to configuration (1).

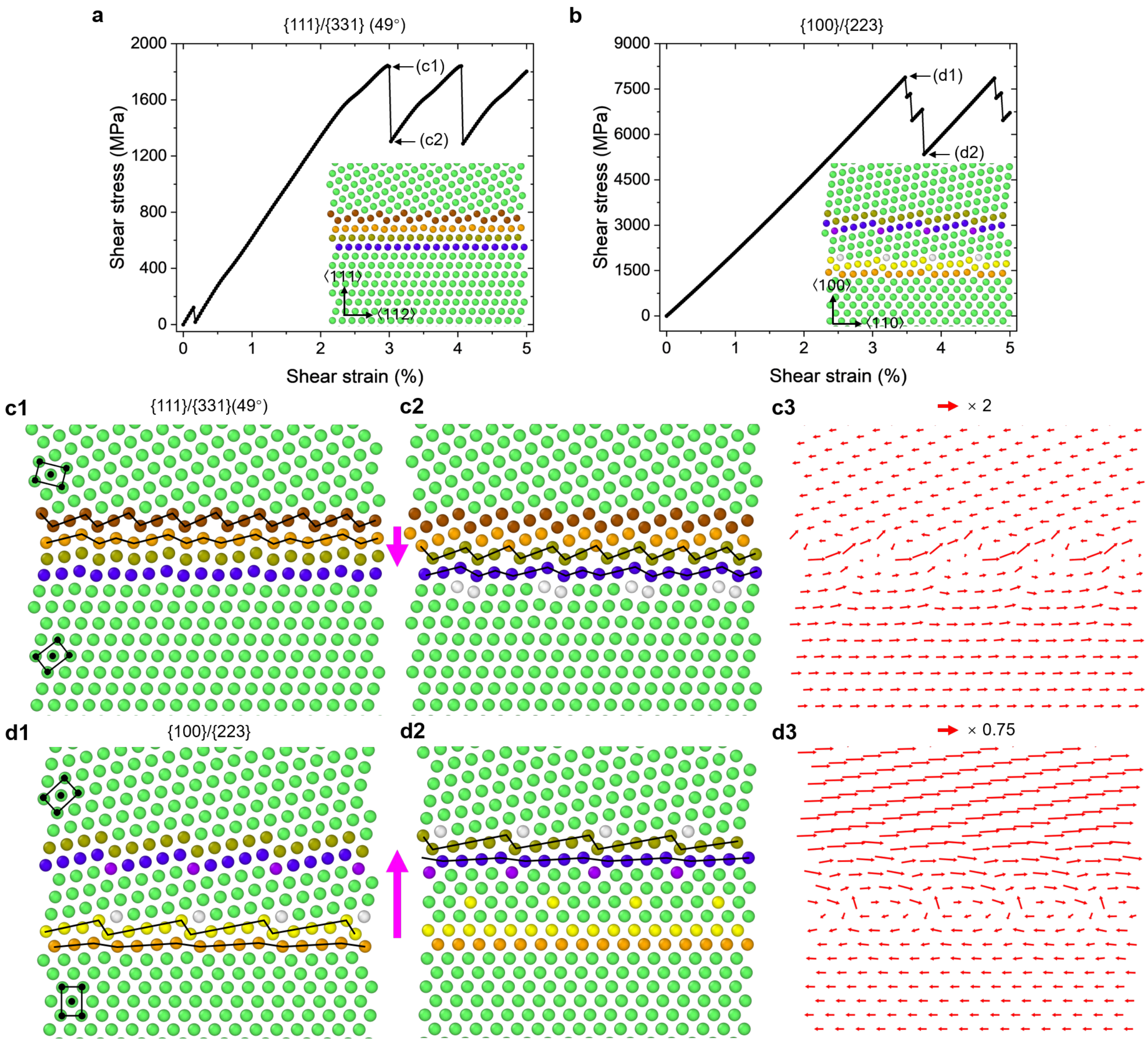

**Fig. S12 | Spatially periodic atomic shuffling during shear-coupled GB migration.** Shear stress as a function of applied shear strain for **a,** {111}/{331}(49°) GB loaded in direction 2 and **b,** {100}/{223} GB. Insets show the corresponding initial GB structures. Atomic configurations of the {111}/{331}(49°) GB, **c1,** before and, **c2,** after the selected migration interval. **d1-d2,** Corresponding atomic configurations of the {100}/{223} GB. Selected atomic sites are shown in distinct colors to facilitate structural tracking. Black lines trace the terminating atomic planes, outlined crystallographic cells indicate the orientations of the two grains, and magenta arrows mark the directions of GB migration. **c3, d3,** Differential atomic displacement vectors from **c1** to **c2** and from **d1** to **d2**, respectively. The red vectors are scaled by factors of 2 and 0.75, respectively. All atomic configurations are viewed along the ⟨110⟩ tilt axis.

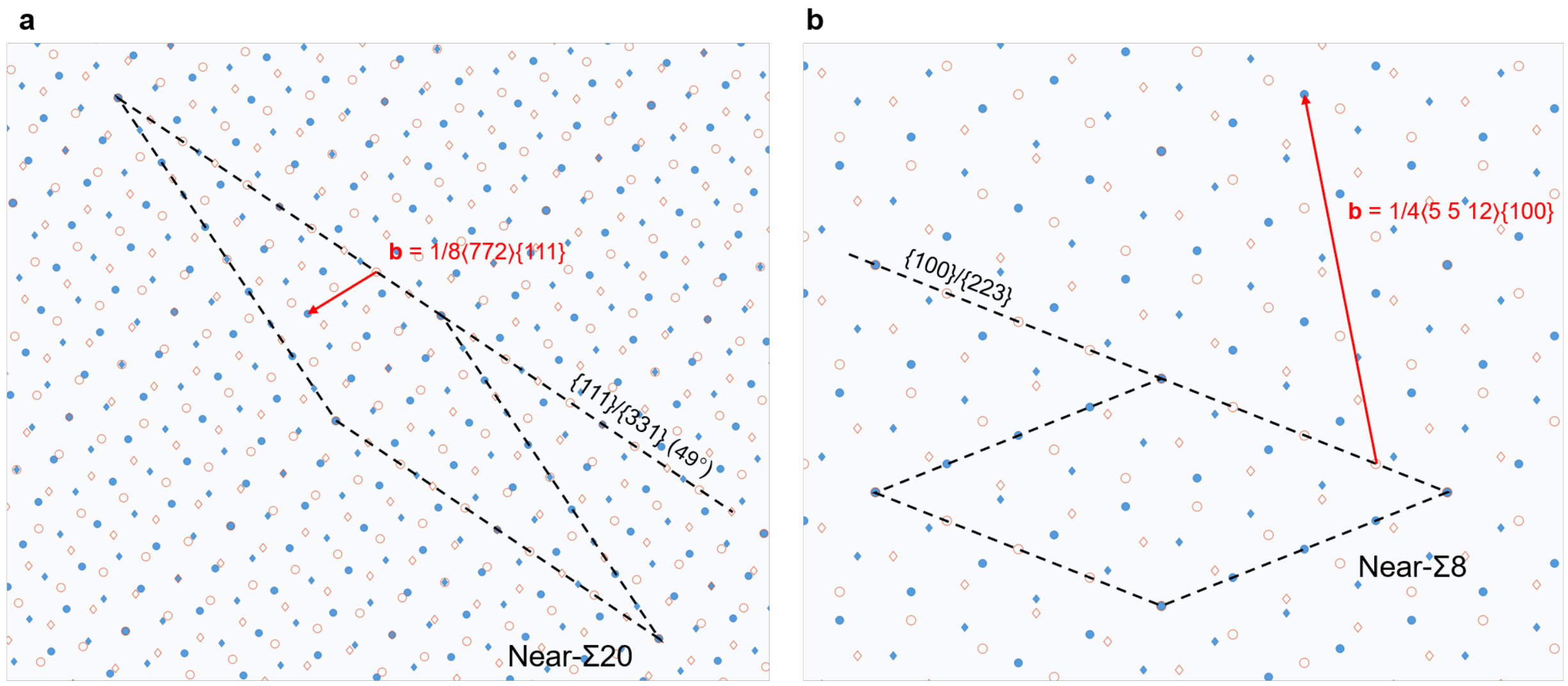

**Fig. S13 | Analysis of near-CSL dichromatic patterns associated with shear-coupled migration.** **a**, {111}/{331}(49°) GB. **b**, {100}/{223} GB. Colors distinguish the grain lattices, and symbol shapes distinguish layers along the ⟨110⟩ tilt axis. Dashed lines indicate the near-CSL cells and GB traces. Red arrows mark DSC vectors selected by comparison with the signed tangential-to-normal displacement ratios measured during migration (Fig. S12). Crystallographic directions are referenced to the grains indicated in each panel.

## Supplementary tables

### Table S1 | Hierarchical memory layers and access policies in GBClaw.

| Level | Layer | Category | Purpose | Storage | Agent access | Update |
|---|---|---|---|---|---|---|
| 0 | Domain knowledge | Procedural memory | Encode domain-specific workflows, tool contracts, and task execution rules | Built into tools and instructions | Implicitly accessed through tool invocation | Manual update only |
| 1 | Project summary | Working memory | Give the orchestrator bounded project intent and conclusions for planning the next step | Working context and database file | Always in orchestrator context | Incrementally updated |
| 2 | Conversation state | Episodic memory | Keep track of recent communication, decisions, and unresolved tasks | Working context and database file | Always in orchestrator context | Periodically compressed and refreshed |
| 3 | Facts/Runs/Conclusions | Semantic memory | Store durable facts, conclusions, run summaries, and pointers to artifacts | Database file | On-demand retrieval | Updated with provenance |
| 4 | Artifacts | External grounding | Store raw outputs, figures, trajectories, logs, code, and other files for traceable evidence | File system in artifact directory | On-demand retrieval | Persistent unless manually deleted by user |
| 5 | Full Chat/work history | Human reference | Preserve the full append-only record for audit and human review | Database file | Excluded from agent retrieval | Append-only |

**Table S2 | Reference Σ GBs with ⟨110⟩ tilt axis and low-index near-Σ GBs for Σ < 50.** STGB: symmetric tilt GB; ATGB: asymmetric tilt GB. $\theta_\Sigma$ and $\theta$ denote the misorientation angles of Σ and near-Σ GBs, respectively; $\Delta\theta$ denotes their difference.

| Exact Σ GBs | | | | | Near-Σ GBs | | | |
|---|---|---|---|---|---|---|---|---|
| Σ | $\theta_\Sigma$ (°) | GB plane 1 | GB plane 2 | Character | Approx. plane 2 | $\Delta\theta$ (°) | $\theta$ (°) | GB plane pair |
| 3 | 70.53 | (1 -1 1) | (1 -1 -1) | STGB | | | | |
| | 70.53 | (0 0 -1) | (-2 2 -1) | ATGB | | | | |
| | 70.53 | (-1 1 0) | (-1 1 4) | ATGB | | | | |
| | 70.53 | (1 -1 -1) | (-1 1 -5) | ATGB | | | | |
| 9 | 38.94 | (-1 1 -1) | (-1 1 -5) | ATGB | | | | |
| | 38.94 | (1 -1 0) | (7 -7 8) | ATGB | (1 −1 1) | 3.68 | 35.26 | {111}/{110} |
| | 38.94 | (-1 1 -1) | (-11 11 1) | ATGB | (1 −1 0) | 3.68 | 35.26 | |
| | 38.94 | (0 0 -1) | (4 -4 -7) | ATGB | (1 −1 −2) | 3.68 | 35.26 | {100}/{112} |
| | | | | | (2 −2 −3) | 4.37 | 43.31 | {100}/{223} |
| 11 | 50.48 | (0 0 -1) | (-6 6 -7) | ATGB | (−1 1 −1) | 4.26 | 54.74 | {111}/{100} |
| | 50.48 | (-1 1 -1) | (-1 1 -19) | ATGB | (0 0 -1) | 4.26 | 54.74 | |
| | 50.48 | (1 -1 0) | (7 -7 12) | ATGB | (1 -1 2) | 4.26 | 54.74 | {110}/{112} |
| | | | | | (2 −2 3) | 3.79 | 46.69 | {110}/{223} |
| 17 | 86.63 | (0 0 -1) | (12 -12 -1) | ATGB | (1 −1 0) | 3.37 | 90.00 | {100}/{110} |
| | 86.63 | (1 -1 0) | (1 -1 24) | ATGB | (0 0 1) | 3.37 | 90.00 | |
| 19 | 26.53 | (-1 1 0) | (-17 17 12) | ATGB | (−3 3 2) | 1.29 | 25.24 | {110}/{332} |
| | 26.53 | (0 0 -1) | (-6 6 -17) | ATGB | (−1 1 −3) | 1.29 | 25.24 | {100}/{113} |
| 27 | 31.59 | (1 -1 -1) | (11 -11 -1) | ATGB | (1 −1 0) | 3.68 | 35.26 | {111}/{110} |
| | 31.59 | (0 0 1) | (10 -10 23) | ATGB | (1 −1 2) | 3.68 | 35.26 | {100}/{112} |
| 33a | 20.05 | (1 -1 -1) | (13 -13 -5) | ATGB | (3 −3 −1) | 1.95 | 22.00 | {111}/{331} |
| | | | | | (2 −2 −1) | 4.26 | 15.79 | {111}/{221} |
| 33c | 58.99 | (-1 1 -1) | (1 -1 -19) | ATGB | (0 0 -1) | 4.26 | 54.74 | {111}/{100} |
| | 58.99 | (0 0 -1) | (-20 20 -17) | ATGB | (-1 1 -1) | 4.26 | 54.74 | |
| 43 | 80.63 | (0 0 -1) | (30 -30 -7) | ATGB | (3 −3 −1) | 3.89 | 76.74 | {100}/{331} |

**Table S3 | Determination of Burgers vectors $\mathbf{b}_\mathrm{s}$ of SGBMDs in Fig. 3.**

| GB pairs | $N_\mathrm{u}$ | $N_\mathrm{l}$ | $n_\mathrm{u}$ | $n_\mathrm{l}$ | $\delta$ | $\Delta N$ | $\mathbf{d}_\mathrm{l}/a_0$* | $\mathbf{b}_\mathrm{s}/a_0$* |
|---|---|---|---|---|---|---|---|---|
| {111}/{100} | 7 | 8 | 5 | 6 | $\frac{2}{35}$ | $\frac{2}{7}$ | $\frac{1}{4}\langle 112\rangle\{111\}$ | $\frac{1}{14}\langle 112\rangle\{111\}$ |
| {111}/{110} | 11 | 18 | 5 | 8 | $-\frac{2}{55}$ | $-\frac{2}{11}$ | $\frac{1}{4}\langle 112\rangle\{111\}$ | $\frac{1}{22}\langle 112\rangle\{111\}$ |
| {111}/{331}(22°) | 8 | 10 | 2 | 3 | $\frac{1}{4}$ | $\frac{1}{2}$ | $\frac{1}{4}\langle 112\rangle\{111\}$ | $\frac{1}{8}\langle 112\rangle\{111\}$ |
| {100}/{110} | 12 | 17 | 5 | 7 | $-\frac{1}{60}$ | $-\frac{1}{12}$ | $\frac{1}{2}\langle 110\rangle\{100\}$ | $\frac{1}{24}\langle 110\rangle\{100\}$ |

* $a_0$ denotes the lattice constant. Vector coefficients are reported without signs, and the terminating plane specifies the reference grain. Here $\delta$ is the difference between the lower-to-upper interval ratios in the core and reference spans, and $\Delta N$ is the accumulated mismatch defined in Note 2. Positive and negative $\Delta N$ indicate an excess and a deficit of lower-grain row intervals relative to the reference ratio, respectively.

**Table S4 | Grain orientations and dimensions of compact periodic supercells for 14 selected general asymmetric GBs.** The *x*-direction follows the applied shear direction, the *y*-direction is normal to the GB, and the *z*-direction follows the common ⟨110⟩ tilt axis. Primary lattice repeats $\boldsymbol{\lambda}$, cell sizes and residual lattice mismatches are listed. Crystallographic repeat vectors are expressed in units of $\boldsymbol{a_0}$ and referenced to the corresponding grain orientations.

| GB plane pairs | Orientation of grain 1 $(\boldsymbol{x}, \boldsymbol{y}, z)$ | Orientation of grain 2 $(\boldsymbol{x}, \boldsymbol{y}, z)$ | $\boldsymbol{\lambda}$ (Grain 1, grain 2) | Supercell dimensions* (nm³) | Lattice mismatch |
|---|---|---|---|---|---|
| {100}/{110} | $[00\bar{1}], [\bar{1}10], [110]$ | $[1\bar{1}0], [00\bar{1}], [110]$ | $12[00\bar{1}], \quad \frac{17}{2}[1\bar{1}0]$ | 8.45×23.95×1.00 | 0.17% |
| {100}/{112} | $[\bar{1}\bar{1}1], [112], [\bar{1}10]$ | $[\bar{1}\bar{1}0], [001], [\bar{1}10]$ | $\frac{9}{2}[\bar{1}\bar{1}1], \quad \frac{11}{2}[\bar{1}\bar{1}0]$ | 2.74×24.60×1.00 | 0.21% |
| {100}/{113} | $[2\bar{3}\bar{3}], [311], [0\bar{1}1]$ | $[0\bar{1}\bar{1}], [100], [0\bar{1}1]$ | $\frac{3}{2}[2\bar{3}\bar{3}], \quad 5[0\bar{1}\bar{1}]$ | 2.48×24.97×1.00 | 0.50% |
| {100}/{223} | $[4\bar{3}\bar{3}], [322], [0\bar{1}1]$ | $[0\bar{1}\bar{1}], [100], [0\bar{1}1]$ | $\frac{1}{2}[4\bar{3}\bar{3}], \quad 2[0\bar{1}\bar{1}]$ | 1.01×24.67×1.00 | 2.99% |
| {100}/{331} | $[6\bar{1}\bar{1}], [133], [0\bar{1}1]$ | $[0\bar{1}\bar{1}], [100], [0\bar{1}1]$ | $\frac{3}{2}[6\bar{1}\bar{1}], \quad \frac{13}{2}[0\bar{1}\bar{1}]$ | 3.24×25.13×1.00 | 0.59% |
| {111}/{221}(16°) | $[\bar{1}\bar{1}4], [221], [\bar{1}10]$ | $[\bar{1}\bar{1}2], [111], [\bar{1}10]$ | $2[\bar{1}\bar{1}4], \quad \frac{7}{2}[\bar{1}\bar{1}2]$ | 3.00×25.05×1.00 | 1.03% |
| {111}/{221}(55°) | $[114], [22\bar{1}], [\bar{1}10]$ | $[\bar{1}\bar{1}2], [111], [\bar{1}10]$ | $2[114], \quad \frac{7}{2}[\bar{1}\bar{1}2]$ | 3.00×25.05×1.00 | 1.03% |
| {111}/{331}(22°) | $[\bar{1}\bar{1}6], [331], [\bar{1}10]$ | $[\bar{1}\bar{1}2], [111], [\bar{1}10]$ | $[\bar{1}\bar{1}6], \quad \frac{5}{2}[\bar{1}\bar{1}2]$ | 2.16×25.24×1.00 | 0.66% |
| {111}/{331}(49°) | $[116], [33\bar{1}], [\bar{1}10]$ | $[\bar{1}\bar{1}2], [111], [\bar{1}10]$ | $[116], \quad \frac{5}{2}[\bar{1}\bar{1}2]$ | 2.16×25.24×1.00 | 0.66% |
| {111}/{100} | $[01\bar{1}], [\bar{1}00], [011]$ | $[\bar{2}1\bar{1}], [\bar{1}\bar{1}1], [011]$ | $\frac{7}{2}[01\bar{1}], \quad 2[\bar{2}1\bar{1}]$ | 1.73×23.88×1.00 | 1.03% |
| {111}/{110} | $[00\bar{1}], [\bar{1}10], [110]$ | $[\bar{1}1\bar{2}], [\bar{1}11], [110]$ | $11[00\bar{1}], \quad \frac{9}{2}[\bar{1}1\bar{2}]$ | 3.87×24.43×1.00 | 0.21% |
| {110}/{112} | $[11\bar{1}], [112], [1\bar{1}0]$ | $[00\bar{1}], [110], [1\bar{1}0]$ | $\frac{7}{2}[11\bar{1}], \quad 6[00\bar{1}]$ | 2.13×24.88×1.00 | 1.03% |
| {110}/{223} | $[33\bar{4}], [223], [1\bar{1}0]$ | $[00\bar{1}], [110], [1\bar{1}0]$ | $[33\bar{4}], \quad 6[00\bar{1}]$ | 2.07×24.80×1.00 | 2.82% |
| {110}/{332} | $[11\bar{3}], [332], [1\bar{1}0]$ | $[00\bar{1}], [110], [1\bar{1}0]$ | $3[11\bar{3}], \quad 10[00\bar{1}]$ | 3.51×26.15×1.00 | 0.50% |

* The dimension in the *y*-direction includes a vacuum layer of ~4 nm.

**Table S5 | Tools available to the orchestrator and specialized agents in GBClaw.**

| Agents | Tool name(s) | Usage |
|---|---|---|
| Orchestrator | list_available_potentials | Identify interatomic interaction models available for simulations. |
| | list_coding_environments | Determine which execution environments are available for custom code. |
| | search_web | Collect online scientific information relevant to the investigation. |
| | delegate_*(subagents) | Route research tasks to the appropriate specialist agents. |
| | get_project_plan<br>get_project_plan_draft | Recover the current research workflow and its pending draft. |
| | approve_project_plan_draft<br>reject_project_plan_draft | Accept or reject a proposed workflow before implementation. |
| | update_project_plan_step | Record the status of individual workflow steps. |
| | report_artifact_usage | Report the space occupied by stored computational artifacts. |
| Planning | create_project_plan_draft | Translate research objectives into an ordered, reviewable task plan. |
| | get_project_plan<br>revise_project_plan_draft | Recover the current plan and revise proposed research steps. |
| Structure | analyze_tilt_geometry | Establish the common tilt axis and relative grain rotation. |
| | compare_miller_planes | Assess the relative orientations of specified crystallographic planes. |
| | match_csl_tilt_references | Associate a boundary geometry with nearby exact CSL misorientations. |
| | score_csl_tilt_plane_pairs | Assess candidate CSL terminating-plane pairs for selection. |
| | suggest_low_index_near_csl_planes | Find low-index terminations close to a reference CSL geometry. |
| | list_csl_sigmas | List available CSL misorientations and their rotation branches. |

| Agents | Tool name(s) | Usage |
|---|---|---|
| | list_csl_gb_planes<br>list_csl_gb_planes_batch | Obtain terminating-plane pairs for individual or multiple CSL references. |
| | get_tilt_gb_orientation | Define grain coordinate systems and compatible lattice repeat lengths. |
| | get_dichromatic_map_options<br>render_dichromatic_map | Set visualization options and display superposed grain lattices. |
| | prepare_dichromatic_cell_selection<br>get_dichromatic_cell_selection | Prepare and recover a user-selected dichromatic unit cell. |
| | apply_dichromatic_cell_strain | Determine the strain required to match a selected near-CSL cell. |
| | analyze_dichromatic_cell_edges | Determine displacement vectors associated with selected cell edges. |
| | query_csl_orientation | Recover previously stored CSL crystallographic data. |
| LAMMPS | list_available_potentials | Identify interatomic interaction models available for simulations. |
| | convert_cfg_to_lammps | Prepare LAMMPS input structures from CFG atomic configurations. |
| | infer_tilt_orientation | Establish grain orientations from specified boundary-plane pairs. |
| | estimate_gb_box_size | Estimate simulation-cell dimensions for bicrystal construction. |
| | write_gb_minimize_config<br>run_gb_minimize | Set construction parameters and obtain relaxed bicrystal configurations. |
| | write_structure_minimize_config<br>run_structure_minimize | Set relaxation parameters and minimize an existing atomic configuration. |
| | write_gb_energy_config<br>run_gb_energy | Set calculation parameters and evaluate interfacial excess energy. |
| | write_gb_shear_config<br>run_gb_shear | Define loading conditions and calculate the boundary shear response. |
| | write_gb_neb_config<br>run_gb_neb | Define transition endpoints and determine minimum energy paths. |

| Agents | Tool name(s) | Usage |
|---|---|---|
| Analysis | analyze_gb_periodicity | Measure structural repeat lengths along the grain boundary. |
| | stress_strain_analysis | Locate stress drops and associated deformation events in shear histories. |
| | plot_stress_strain_comparison | Display shear responses together for direct comparison. |
| | analyze_neb_mep<br>plot_neb_mep | Extract transition barriers and visualize the calculated energy path. |
| | render_atomic_structure | Produce atomic-structure images through OVITO. |
| | render_gb_plan_view | Display the interface along its normal for structural inspection. |
| Coding | ls; glob; grep | Find relevant files and locate information within their contents. |
| | read_file; write_file; edit_file; delete | Inspect, create, modify or remove files within the coding sandbox. |
| | execute | Run custom commands within the coding sandbox. |
| | publish_coding_result | Validate coding outputs and make them available to the workflow. |
| All agents | query_project_memory<br>search_project_memory | Find project records relevant to the current task. |
| | read_agent_conclusion<br>get_task_state<br>read_tool_output | Recover agent findings, task progress and complete tool results. |
| All except Coding | list_run_artifacts<br>read_run_artifact | Find output files associated with a run and inspect their contents. |

## Note 1. Calculation of GB energy

GB energies were calculated using periodic supercells containing a bicrystal with one internal GB and two free surfaces. For each constituent grain, a periodic bulk reference and a slab reference containing two free surfaces were constructed with the same orientation and relaxed in-plane dimensions as the bicrystal. Atomic coordinates and the cell length normal to the GB were relaxed while the in-plane dimensions remained fixed. The reference bulk energy per atom therefore retained the elastic contribution associated with the strain imposed by in-plane periodicity. The GB excess energy was calculated as

$$\gamma_{\mathrm{GB}} = \frac{E_{\mathrm{bi}} - N_{\mathrm{t}} e_{\mathrm{t}}^{\mathrm{bulk}} - N_{\mathrm{b}} e_{\mathrm{b}}^{\mathrm{bulk}}}{A} - \gamma_{\mathrm{t}}^{\mathrm{s}} - \gamma_{\mathrm{b}}^{\mathrm{s}} \tag{S1}$$

Here $E_{\mathrm{bi}}$ is the relaxed bicrystal energy, $N_{\mathrm{t}}$ and $N_{\mathrm{b}}$ are the numbers of atoms in the top and bottom grains, and $e_{\mathrm{t}}^{\mathrm{bulk}}$ and $e_{\mathrm{b}}^{\mathrm{bulk}}$ are the corresponding strained bulk energies per atom. $A$ is the interfacial area, equal to the area of each free surface. The surface energy of each grain $i$ was calculated as

$$\gamma_i^{\mathrm{s}} = \frac{E_i^{\mathrm{slab}} - N_i^{\mathrm{slab}} e_i^{\mathrm{bulk}}}{2A} \tag{S2}$$

Here $E_i^{\mathrm{slab}}$ and $N_i^{\mathrm{slab}}$ are the total energy and number of atoms in the two-surface slab, and $e_i^{\mathrm{bulk}}$ is the strained bulk energy per atom. Equation (S1) isolates the GB excess energy by subtracting both the strain-matched bulk energies and the contributions from both external free surfaces from the total bicrystal energy.

**Note 2. Identification of SGBMDs from atomistic simulations and HRTEM images**

SGBMDs can be identified from projected atomic-column arrangements in simulated GB structures. This analysis provides a basis for interpreting experimental high-resolution transmission electron microscopy (HRTEM) images and evaluating their consistency with simulated defect structures. Atomistic simulations provide three-dimensional atomic coordinates, whereas HRTEM images acquired along a zone axis resolve projected atomic columns in the two-dimensional image plane. Because a single projection does not resolve atomic positions along the beam direction, the atomic volumetric strain used to highlight dislocation cores in Fig. 2 and Fig. S5 cannot be evaluated directly from these images. We therefore describe a procedure for identifying SGBMDs and determining their Burgers vectors through analysis of projected atomic-column arrangements.

For the four asymmetric, incommensurate GBs, $\{111\}/\{100\}$, $\{111\}/\{110\}$, $\{111\}/\{331\}$(22°), and $\{100\}/\{110\}$, we relaxed extended bicrystal models along the horizontal direction, with free surfaces at both ends. These boundary conditions allowed SGBMDs to form during relaxation, producing the structures shown in Fig. 4a-d. We then traced close-packed $\{111\}$ planes inclined to the GB plane, as indicated by the white solid lines. The orientations of the terminating planes determine the inclination angles of these $\{111\}$ planes and the spacing between their intersections with the interface. When these spacings differ between the adjoining grains, more $\{111\}$ planes intersect a given boundary length on the side with the smaller spacing. The resulting extra $\{111\}$ half-planes terminate at the interface. Yellow "⊥" symbols mark these terminations, identifying the PGBMD cores[22]. The shortest complete repeat of the PGBMD arrangement defines the local GB structural period λ, indicated by the red double-headed arrows. Crystallographic equivalence between successive repeats was verified from the full atomic registry of the simulated structure.

Localized interruptions in the periodic PGBMD arrangement identify the projected SGBMD cores, whose spans are indicated by the green double-headed arrows in Fig. 4a-d. To determine the Burgers vector of an SGBMD, we quantified the residual mismatch within its core span relative to the periodic PGBMD arrangement in a neighboring reference span of length λ. Let $N_\mathrm{l}$ and $N_\mathrm{u}$ denote the numbers of projected intervals in the lower and upper grains within this reference span, respectively. Each interval lies between successive intersections of the inclined close-packed planes, here $\{111\}$, with the GB. The reference ratio is defined as $\widetilde{N} = N_\mathrm{l}/N_\mathrm{u}$ . For an SGBMD core span containing $n_\mathrm{u}$ intervals in the upper grain and $n_\mathrm{l}$ intervals in the lower grain, the accumulated mismatch is $\Delta N_\mathrm{l} = n_\mathrm{l} - n_\mathrm{u}\widetilde{N}$. Positive and negative values of $\Delta N_\mathrm{l}$ indicate an excess and a deficit, respectively, of lower-grain intervals relative to the reference ratio. The Burgers vector is then calculated as $\mathbf{b} = \mathbf{d_l}\,|\Delta N_\mathrm{l}|$, where $\mathbf{d_l}$ denotes the row translation vector of the lower grain along the horizontal direction. Its magnitude equals the spacing between successive intersections of the inclined $\{111\}$ planes with the GB. For the SGBMDs examined here, the Burgers vectors lie within the GB plane and perpendicular to the tilt axis, so their full magnitudes are resolved in this projection. Atomic registry along the tilt axis was checked separately using the near-CSL patterns in Fig. S7. Table S3 summarizes the reference and core interval counts, signed mismatches, and corresponding Burgers vectors for all four GBs.

Atomic volumetric-strain maps provide an additional check of the geometrically identified SGBMDs (Fig. S5). Local dilatational distortions produce strain contrast around the dislocation cores, reflecting variations in atomic spacing. The localized features in Fig. S5b to d support the SGBMD assignments in the corresponding GBs. In contrast, the strain map of the $\{111\}/\{100\}$ GB does not resolve individual SGBMD cores (Fig. S5a). This lack of strain localization is due to core spreading, consistent with the low sliding resistance. The geometrical analysis in Fig. 4 identifies the associated changes in atomic-row correspondence. These strain maps therefore provide complementary simulation evidence for the SGBMD assignments, while experimental identification relies on projected atomic-column geometry

Fig. S6 illustrates the identification of SGBMDs from projected atomic-column arrangements in *in situ* HRTEM images of a $\{111\}/\{331\}(22°)$ GB undergoing sliding. Each image in Fig. S6a, b shows two PGBMDs. Solid white lines trace extra $\{111\}$ half-planes, whose interfacial terminations identify the PGBMD cores. In Fig. S6c, an additional $\{111\}$ half-plane identifies an SGBMD. The associated atomic-column arrangement is consistent with the simulated structure shown in Fig. 4c. This agreement supports the application of the procedure described in this Note to identify SGBMDs and determine their Burgers vectors from projected atomic-column arrangements in HRTEM images.

## Note 3. SGBMD character from near-CSL analysis

Near-CSL dichromatic patterns (Fig. S7) provided an independent crystallographic check of the Burgers vectors determined in Note 2. To construct these patterns, the two grain lattices were superimposed in their prescribed orientations using the structural period determined from atomistic simulation. Then, biaxial elastic strains and a relative lattice rotation were solved to produce a common periodic cell with an exact CSL. In Fig. S7, colors distinguish the two grain lattices, while circles and diamonds distinguish atomic layers along the ⟨110⟩ projection direction. Within each GB plane, the shortest nonzero vector connecting lattice sites of the two grains was identified as the minimum DSC translation within the same atomic layer. This translation connects crystallographically equivalent interfacial states and corresponds to the Burgers vector $\mathbf{b}_s$ of the SGBMD in the corresponding GB.

In each panel of Fig. S7, the black arrow inside the solid box denotes the shortest DSC vector and thus the Burgers vector for the corresponding GB. Expressed in units of the lattice constant $a_0$, this vector is $1/14\langle112\rangle$ for {111}/{100}, $1/22\langle112\rangle$ for {111}/{110}, $1/8\langle112\rangle$ for {111}/{331}(22°), and $1/24\langle110\rangle$ for {100}/{110}. The ⟨112⟩ and ⟨110⟩ directions are defined relative to the grains terminated by {111} and {100} planes, respectively. These Burgers-vector assignments agree with the projected atomic-column analysis in Note 2. In addition, each dashed circle identifies lattice-site pairs defining the shortest DSC vectors opposite to that shown in the corresponding solid box. These antiparallel DSC vectors have equal magnitudes and represent GB sliding in opposite senses.

The 22° and 49° variants of the {111}/{331} GB (Figs. S7c and S9a) illustrate the respective DSC vectors and demonstrate how atomic configurations govern direction-dependent GB shearing responses. Both variants share the shortest DSC vector and thus Burgers vector $\mathbf{b}_s$ = 1/8⟨112⟩{111} and a primary period $\lambda$ = 2.16 nm. For the {111}/{331}(49°) GB (Fig. S9b), red and green arrows denote +$\mathbf{b}_s$ and -$\mathbf{b}_s$, respectively. Black arrows indicate resistance from second-nearest-neighbor atoms within each structural period beneath the terminating plane. Because this intraplanar resistance opposes sliding along $\mathbf{b}_s$ more strongly than along -$\mathbf{b}_s$, -$\mathbf{b}_s$ represents the favored sliding mode; conversely, sliding along $\mathbf{b}_s$ is unfavored, which promotes shear-coupled migration for the 49° variant. In contrast, for the 22° variant, sliding is favored along both +$\mathbf{b}_s$ and -$\mathbf{b}_s$, due to similar resistance encountered in both directions.

## Note 4. Nonuniform glide of SGBMDs during pure sliding

Shear simulations of GBs containing multiple SGBMDs were performed while imposing PBCs within the boundary plane. Each supercell contained an array of SGBMDs whose Burgers vectors summed to a lattice repeat along the boundary, resulting in the formation of an extra half-plane. This construction maintained crystallographically equivalent interfacial registry across periodic images. The secondary periodicity vector $\boldsymbol{\Lambda}$ connects neighboring equivalent SGBMDs, and its average magnitude $|\overline{\boldsymbol{\Lambda}}|$ is listed in Table 1. The horizontal supercell length therefore spans multiple secondary periods.

For the $\{111\}/\{331\}$(22°) GB, sliding proceeded through intermittent glide of individual SGBMDs or subsets of the array, punctuated by collective events. Fig. S8 correlates the shear stress response with successive dislocation positions during a representative sliding cycle. Compared with the uniform-sliding response of the compact supercell in Fig. 3c2, this response exhibits finer fluctuations associated with the advance of one or several SGBMDs by integer multiples of the primary structural period λ. Between states (1) and (2), the two central SGBMDs advance while the outer two remain near their previous positions (Fig. S8c1, c2). Continued loading from state (2) to (3) raises the shear stress and produces accumulated dislocation motion along the boundary (Fig. S8c3). This longer interval includes successive local advances during reloading. The smaller stress decrease from state (3) to (4) accompanies further advance of three of the four tracked dislocations, while the second dislocation from the left remains near its previous position (Fig. S8c3, c4). The pronounced drop from state (4) to (5) corresponds to collective glide of all four dislocations (Fig. S8c4, c5), releasing accumulated elastic shear stress during relaxation at fixed imposed strain.

## Note 5. Shear asymmetry of the {111}/{331}(49°) GB

The analysis agent compared the 22° and 49° misorientation variants of the {111}/{331} GB, providing mechanistic insights into shear asymmetry and deformation-mode selection. The two variants arise from two {111} planes whose inclination angles differ by 71° about the ⟨110⟩ tilt axis. Both variants share a primary periodicity $\boldsymbol{\lambda} = 5/2\langle 112\rangle\{111\}$ and an SGBMD Burgers vector $\mathbf{b}_s$ = 1/8⟨112⟩{111} (Table 1, Table S4 and Supplementary Fig. S9). When sheared along direction 1 (the mode of lower shear strength), the 49° variant exhibited a pure-sliding response similar to that of the 22° variant. Its repeated stress drops produced a comparable sawtooth profile, albeit at different stress levels (Figs. 3c2 and Fig. S10a). Atomic configurations and displacement vectors likewise confirmed tangential grain translation without GB migration (Figs. 3c3, c4 and Fig. S10c1-c3). When shear was applied along direction 2, however, the 49° GB underwent shear-coupled migration, whereas the 22° GB maintained pure sliding. Specifically, stress-drop events in the 49° GB coincided with downward boundary motion through the transformation of atomic columns into the opposing grain orientation and coordinated shuffling across two atomic layers (Fig. S10b, d1-d3). Dichromatic mapping based on near-CSL constructions revealed different lattice registries for the two variants (Supplementary Figs. S7c and S9a), suggesting that differences in subsurface stacking sequences and second-nearest-neighbor coordination likely underlie the distinct shear pathways. Supplementary Notes 7 and 8 detail the corresponding shear-coupled migration pathways and crystallographic mapping, respectively. This comparison further illustrates how GBClaw bridges crystallographic characterization with the local atomic mechanisms governing shear responses.

## Note 6. Cyclic loading-unloading and NEB calculations of the {111}/{110} GB

Dynamic replanning enabled GBClaw to leverage researcher guidance to investigate an anomalous shear response and resolve its structural origin. During the initial shear simulation of the {111}/{110} GB, an unusually high initial stress peak was followed by a sharp drop and steady-state sliding at a substantially lower stress. In response to a researcher's inquiry, the orchestrator dynamically expanded the simulation plan to execute a loading-unloading cycle and NEB calculations (Fig. S11a). After unloading to zero applied shear strain, the bicrystal retained a nonzero shear stress, whereas the initial configuration was stress free (inset in Fig. S11a). The analysis agent then compared the two zero-strain configurations and mapped the atomic displacements between them (Fig. S11b). This comparison showed that unloading did not restore the initial interfacial structure. The two relaxed configurations instead occupied distinct local minima under identical zero-strain constraints.

To characterize the energy landscape connecting these states, GBClaw performed an NEB calculation using the relaxed initial and reconstructed GB configurations as endpoints at zero imposed shear strain. The resulting minimum energy path exhibited an energy barrier of approximately $20\ \mathrm{meV}/a_0$, where $a_0$ = 3.52 Å is the lattice constant, and terminated at a state with lower energy than the initial configuration (Fig. S11c). This result indicated the metastability of the initial GB structure. Together with the localized atomic displacements at the saddle-point state relative to the initial state (Fig. S11d), these results revealed a loading-induced local reconstruction of the {110} terminating plane against the opposing {111} plane. These findings attributed the initially high sliding resistance to a metastable local structure formed during bicrystal construction and relaxation. Reloading the reconstructed zero-strain configuration did not reproduce the anomaly and instead yielded the normal stress-strain response shown in Fig. 3b2. This investigation demonstrates how human-in-the-loop replanning empowers GBClaw to unravel unexpected simulation phenomena and pinpoint their atomistic reconstruction mechanisms.

The {111}/{110} GB was loaded through 70 increments of $\Delta\gamma = 2.5 \times 10^{-4}$ to a shear strain of 1.75%, and then unloaded through 70 increments of the same magnitude to zero applied strain (Fig. S11a). Although the initial configuration was stress free, the unloaded configuration retained a nonzero shear stress. This configuration was subsequently reloaded through 200 increments of $\Delta\gamma = 0.5 \times 10^{-4}$ to a shear strain of 1% (Fig. 3b2). The pronounced initial stress drop occurred only during the first loading sequence and did not recur upon reloading. The residual shear stress and endpoint displacements in Fig. S11a, b show that unloading did not restore the initial interfacial structure. The NEB endpoints compare the relaxed initial and reconstructed states at the same zero applied strain (Fig. S11c). The NEB settings and convergence criterion are specified in Methods. The final force residual reached $2 \times 10^{-8}$ eV $Å^{-1}$, below the convergence tolerance specified there. The energy profile and saddle-point displacement field are shown in Fig. S11c, d.

**Note 7. Shear-coupled migration**

GBClaw evaluated the mechanisms governing the shear-coupled migration mode identified in Fig. 2e using compact supercell models of the {111}/{331}(49°) and {100}/{223} GBs as representative cases. Both exhibited pronounced stress drops associated with discrete migration events (Fig. S12a, b). Comparing atomic configurations immediately before and after selected stress drops linked these events to locally nonuniform atomic shuffling. Both boundaries joined an atomically flat, low-index termination in one grain to a stepped termination in the opposing grain.

For the {111}/{331}(49°) GB, downward migration transformed atomic columns from the flat termination of the lower grain to the stepped configuration of the upper grain (Fig. S12c1, c2). The corresponding displacement field revealed spatially repeated, coordinated shuffling characterized by a localized linear pattern among neighboring atomic columns (Fig. S12c3). Conversely, upward migration of the {100}/{223} GB transformed atomic columns from the stepped termination of the upper grain into the flatter configuration of the lower grain (Fig. S12d1, d2). Its displacement field likewise exhibited spatially periodic but locally nonuniform shuffling (Fig. S12d3), featuring a localized rotational pattern distinct from that of the {111}/{331} GB. In these compact supercells, shear-coupled migration occurred collectively along the boundary rather than through the propagation of isolated GB disconnections. Supplementary Note 8 and Fig. S13 present the corresponding crystallographic mappings derived from dichromatic patterns. These results further demonstrate the utility of near-CSL constructions and dichromatic patterns for determining DSC vectors associated with distinct deformation modes in asymmetric incommensurate GBs.

**Note 8. Determination of DSC vectors in shear-coupled migration**

Using the near-CSL construction described in Note 3, we analyzed the dichromatic patterns to identify the crystallographic mapping associated with shear-coupled migration. Atomic trajectories yielded the signed tangential grain translation $u_t$ and normal GB displacement $u_n$ during migration of the {111}/{331}(49°) and {100}/{223} GBs. Comparing the signed ratios of $u_t/u_n$ with the corresponding component ratios of candidate DSC vectors identified the closest DSC vectors of $1/8\langle 772\rangle$ and $1/4\langle 5\ 5\ 12\rangle$ for each case, referenced to the grains terminated by {111} and {100} planes, respectively.

Circles and diamonds in Fig. S13 distinguish atomic layers at different depths along the common ⟨110⟩ tilt axis. Candidate connections join sites in different grains within corresponding depth layers and may include a displacement normal to the interface. Their components were resolved parallel and perpendicular to the dashed GB trace in the projection plane.

PBCs constrain the elastic state of the simulated bicrystals, while the near-CSL construction also introduces strain to enforce periodic matching. These elastic states can differ, resulting in small differences between the geometrically predicted and simulated displacement ratios. Candidate mapping was therefore selected based on approximate agreement with the signed displacement ratio, correspondence between atomic depth layers, and consistency with the observed structural transformation.